# A front-tracking study of retinal detachment treatment by magnetic drop targeting

Mohammad Amin Amini[1], Gretar Tryggvason[2], and Ehsan Amani[1,*]

[1]*Department of Mechanical Engineering, Amirkabir University of Technology (Tehran Polytechnic), Iran*

[2]*Department of Mechanical Engineering, Johns Hopkins University, Baltimore, MA, 21218, USA*

**Abstract**

We investigate the Ferrofluid Drop Targeting (FDT) for the treatment of the Retinal Detachment (RD), considering, for the first time, the real 3D geometry of an eye and magnets configurations as well as the viscoelastic rheology of the medium, i.e., the Vitreous Humor (VH). A Front-Tracking Method (FTM) is extended to handle a general 3D unstructured Eulerian grid and strong wall effects. The challenges include the accuracy and robustness of the solver when the drop spreads on the retina under the effect of a magnetic field, which necessitates the design of a multi-region Eulerian grid and defining a threshold distance between the front and wall, along with the choice of an effective front smoothing and volume correction FTM sub-algorithms near the walls. After model validations, the effect of different design parameters on important objectives, such as the travel time, settling time, retinal coverage area, and impact compressive stress, are studied. The results reveal that, in addition to the magnetic Bond number, the ratio of the drop-to-VH magnetic permeabilities plays a key role in the terminal shape parameters, like the retinal coverage. Additionally, simultaneously increasing these two parameters, significantly increase the total FDT force, coverage area, and stress concentration, while decreasing the drop-VH surface tension can mitigate the stress concentration on the retina.

**Keywords:** Front-Tracking Method (FTM); Retinal detachment; Ferrofluid drop targeting; Multiphase flow; Viscoelastic fluid

---

[*] Corresponding author. Address: Mechanical Engineering Dept., Amirkabir University of Technology (Tehran Polytechnic), 424 Hafez Avenue, Tehran, P.O.Box: 15875-4413, Iran. Tel: +98 21 64543404. Email: eamani@aut.ac.ir

## Nomenclature

| **Latin symbols** | | | $\rho$ | : | Density, $kgm^{-3}$ |
|---|---|---|---|---|---|
| $A$ | : | Area, $m^2$ | $\tau$ | : | Time scale, $s$ |
| $\boldsymbol{B}$ | : | Magnetic induction vector, $T$ | $\boldsymbol{\tau}$ | : | Stress tensor, $Pa$ |
| $\boldsymbol{D}$ | : | Strain rate tensor, $s^{-1}$ | $\varphi$ | : | Magnetic potential, $A$ |
| $D$ | : | Diameter, $m$ | $\chi$ | : | Magnetic susceptibility, - |
| $\boldsymbol{F}_\sigma$ | : | Surface tension per unit volume, $Nm^{-3}$ | **Subscripts** | | |
| $g$ | : | Gravitational acceleration, $ms^{-2}$ | 0 | : | Initial, total |
| $G_I$ | : | Indicator function gradient, $m^{-1}$ | $c$ | : | Continuous phase, coverage |
| $\boldsymbol{H}$ | : | Magnetic field vector, $Am^{-1}$ | $d$ | : | droplet |
| $H_0$ | : | Characteristic magnetic field, $Am^{-1}$ | $f$ | : | front |
| $I$ | : | Phase-indicator function | $m$ | : | magnetic |
| $L$ | : | Characteristic length, $m$ | mag | : | magnet |
| $\boldsymbol{n}$ | : | Interface normal vector, - | me | : | Magnet to inner eye wall |
| $p$ | : | Pressure, $Pa$ | mm | : | Magnet to magnet |
| $R$ | : | Radius, $m$ | $p$ | : | Polymeric |
| $t$ | : | Time, $s$ | $s$ | : | Solvent |
| $\boldsymbol{u}$ | : | Velocity vector, $ms^{-1}$ | set | : | Settling |
| $U$ | : | Characteristic velocity, $ms^{-1}$ | $t$ | : | Tube |
| $V$ | : | Volume, $m^3$ | tra | : | Traveling |
| $\boldsymbol{x}$ | : | Location, $m$ | $\sigma$ | : | Surface tension |
| **Greek symbols** | | | **Superscripts** | | |
| $\alpha$ | : | Mobility factor, - | * | : | Dimensionless |
| $\delta$ | : | 3D delta function, $m^3$ | **Dimensionless parameters** | | |
| $\sigma$ | : | Surface tension coefficient, $Nm^{-1}$ | $Bo_m$ | : | Magnetic Bond number |
| $\varepsilon$ | : | Threshold distance, $m$ | $Ca$ | : | Capillary number |
| $\zeta$ | : | Sphericity, - | $De$ | : | Deborah number |
| $\eta$ | : | Dynamic viscosity, $Pa.s$ | $La$ | : | Laplace number |
| $\theta$ | : | Angle, degree | $La_m$ | : | Magnetic Laplace number |
| $\lambda$ | : | Relaxation time, $s$ | $Re$ | : | Reynolds number |
| $\mu$ | : | Magnetic permeability, $Hm^{-1}$ | | | |

## 1. Introduction

Retinal Detachment (RD) is characterized by the separation of the neurosensory retina from the underlying retinal pigment epithelium and choroid and is a serious ocular condition that can lead to vision loss. This disease is more prevalent in individuals over the age of 50 and those with certain medical conditions such as myopia, cataract surgery, and diabetic retinopathy [1, 2]. Treatments, aiming to reattach the retina and prevent further vision loss, include pneumatic retinopexy, scleral buckle surgery, vitrectomy, and laser photocoagulation. Park, et al. [3] reviewed surgical techniques for scleral buckling and the complications, such as subretinal hemorrhage and infection, that can arise from invasive procedures.

In a pioneering study, Mefford, et al. [4] experimentally explored RD treatment by the motion of a ferrofluid droplet in the vitreous body, or Vitreous Humor (VH) which is a gel-like viscoelastic fluid filling the space between the retina and lens, under magnetic fields, in order to reattach a detached retina. Here, this novel technique is called Ferrofluid Drop Targeting (FDT).

Subsequently, Afkhami, et al. [5] used the Volume Of Fluid (VOF)- Piecewise Linear Interface Construction (PLIC) method to computationally study FDT. They investigated the deformation and travel time of a ferrofluid droplet through a Newtonian fluid and compared their results with the ones by Mefford, et al. [4]. The effect of non-Newtonian rheology of VH was, however, not accounted for.

Other fields of ophthalmic research have focused on understanding the rheology of VH and its role in the pathogenesis of RD. Modarreszadeh and Abouali [6] proposed a 2-mode Giesekus viscoelastic model, based on the measured complex modulus data by Nickerson, et al. [7], and conducted a numerical study on the mechanical behavior of the human VH under sinusoidal eye motion. They reported sensitivity of the results to the VH constitutive law. In subsequent studies, Bayat et al. [8, 9] performed numerical investigation on the dynamics of partially liquefied VH, a two-phase viscoelastic-Newtonian fluid flow, in a simplified eye geometry, i.e., planar cavity, under an oscillatory motion. They used a 3-mode Giesekus law for VH, calibrated to the measurements by Bonfiglio, et al. [10]. Silva, et al. [11] reviewed the studies on the use of different constitutive laws for VH.

Though the viscoelastic rheology of VH has not yet been taken into consideration in FDT research, the deformation and dynamics of a moving droplet in a viscoelastic matrix in the absence of magnetic fields is well-known. Greco [12] obtained an analytical solution for the fully-developed deformation of a drop in shear flow in the small deformation limit (small capillary numbers). The drop and the matrix were assumed to be quadratic viscoelastic fluids. They compared their predictions with experimental measurements [13] and showed that drop orientation in a viscoelastic matrix is more aligned with the flow, compared to the corresponding Newtonian medium. Khismatullin, et al. [14] performed a 3D computational study of a Newtonian drop in a viscoelastic shear flow, using VOF-Parabolic Reconstruction of Surface Tension (PROST) for capturing the two-phase interface dynamics and comparing the Oldroyd-B and Giesekus rheological models. Verhulst, et al. [15, 16] conducted a computational study using VOF-PROST as well as experiments on drop deformation in shear flow with different Newtonian/viscoelastic drop-matrix combinations. They reported that the drop viscoelasticity has a negligible effect on drop deformation and orientation, while matrix viscoelasticity decreases these factors, with saturation at high Deborah numbers, compared to the Newtonian medium. Habla, et al. [17] numerically investigated the breakup of a Newtonian drop in viscoelastic media using VOF-Multidimensional Universal Limiter for Explicit Solution (MULES) interface capturing, using the Both-Side-Diffusion (BSD) technique to prevent divergence due to the High Weissenberg Number Problem (HWNP). They reported that breakup is hindered by the matrix viscoelasticity. Figueiredo, et al. [18] tested the kernel-conformation tensor approach to resolve HWNP for a series of benchmarks, including a drop in viscoelastic shear flow. For an excellent review of the numerical approaches to tackle HWNP, readers can consult reference [19].

The use of the Front-Tracking Method (FTM) to study drop dynamics in a viscoelastic matrix in the literature is limited to simple geometries and structured grids. Adopting a 3D FTM in a simple shear flow, Mukherjee and Sarkar [20] demonstrated that the matrix viscoelasticity reduces the drop migration velocity away from the walls. Using a 2D FTM for axisymmetric constricted capillary tubes, Muradoglu and coworkers [21, 22] showed that the drop viscoelasticity has significant effect on the drop deformation, especially downstream of the constriction, in contrast to what has been reported for simple shear flows. For capsules, enclosed by a thin membrane, filled with a Newtonain fluid, and transported in a simple shear flow of viscoelastic matrix, Raffiee, et al. [23] reported a morphology which resembles the drop deformation in viscoelastic media. In addition, they pointed out that the matrix viscoelasticity slows down the rotational velocity of the deformed membrane.

Several authors have studied the deformation of a ferrofluid drop under magnetic fields in a Newtonian fluid, e.g., [5, 24, 25]. However, none of them considered the dynamics of a ferrofluid drop in a viscoelastic medium, which is of importance for FDT-RD treatment. In addition, previous computational work on FDT [5] used a simplified 2D geometry. Therefore, the present study aims at filling the gaps in the literature. The novelties of the present work can be summarized as: 1) The use of FDT accounting for the viscoelastic rheology of VH; 2) Considering the real 3D geometry of the eye and magnet configurations for FDT, which requires a multiregion approach; 3) The use of a 3D-unstructured-grid FTM for FDT for a real eye geometry, which can be more accurate than VOF-based approaches provided that proper algorithms are adopted for different steps of FTM on general unstructured Eulerian grids. By analyzing the results, we intend to shed light on the fluid mechanics of FDT and its therapeutic objective parameters, such as travel time, settling time, droplet shape, sphericity, retinal coverage area, and impact compressive stress. To focus on the interaction of the drop and viscoelastic VH, we did not explicitly account for the mechanical properties of the retina layer, which exhibits viscoelastic behavior [26, 27]. This necessitates a fluid-solid interaction framework, which could be studied further. The current study focused on fluid mechanics aspects of FDT, and to link the findings to clinical feasibility, biocompatibility testing [28, 29], involving cytotoxicity, inflammatory response, and long-term retention, of magnetic nano particles used in the drop fluid would be necessary.

The rest of the article is structured as follows: First, the problem definition and governing equations are introduced in section 2. Then, in section 3, the numerical solution procedure is detailed. In section 4.1, the validation of the computational model and particular numerical challenges in applying the FTM to the present problem are addressed. After that, the results of the FDT for a real eye geometry are presented in section 4.2. Finally, the major findings and conclusions of the study are summarized in section 5.

## 2. Mathematical modeling

### 2.1. Problem statement

Figure 1 shows a schematic of the FDT problem. A ferrofluid droplet with an initial spherical shape and a diameter of $D_d = 4\ mm$ is injected through pars plana into the VH of a human eye of an inner diameter of $D_{eye} = 25\ mm$. The droplet is guided towards the damaged retina using a magnetic field created by five cubic magnets of edge length $L_{mag} = 1.5\ mm$. The angle between centers of two adjacent magnets is $\theta_{mm} = 8°$ ($L_{mm} = 2\ mm$), the distance to the inner eye surface is $L_{me} = 1\ mm$, and the pole-to-pole (magnetic) potential difference is $\Delta\phi_0 = 180\ A$. In practice, these magnets can be mounted on the scleral buckle, a piece of band sewn around the eye circumference, e.g., in a scleral buckling procedure [30]. The initial position of the droplet center is at $(x_0, y_0, z_0) = (0, 3, 0)\ mm$.

The study assumes that the ferrofluid droplet is composed of a Newtonian fluid with a density of $\rho_d = 1320\ kg/m^3$, dynamic viscosity of $\eta_d = 80\ Pa.s$, and magnetic permeabilities in the range of $2.002 \times 10^{-6} \leq \mu_d \leq 5.01 \times 10^{-6}\ (H/m)$. The VH is a viscoelastic fluid, modeled by the 3-mode Giesekus model with the rheological properties reported by Bayat, et al. [9] and Bonfiglio, et al. [10], given in table 1. The VH density is $\rho_c = 1000\ kg/m^3$ and its magnetic permeability is $\mu_c = 1.256 \times 10^{-6}\ H/m$, which is assumed equal to the magnetic permeability of vacuum and the medium outside the eye. The surface tension coefficient across the fluids interface is chosen in the range $0.00081 \leq \sigma \leq 0.0135\ (N/m)$.

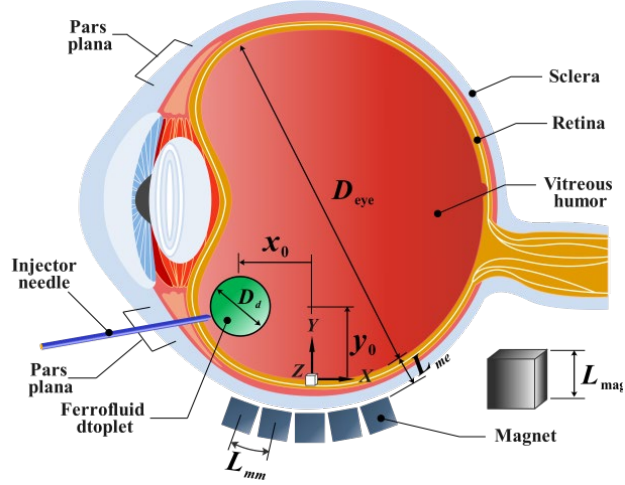

**Figure 1** A schematic of the FDT process and the geometrical parameters.

**Table 1** Parameters of the 3-mode Giesekus viscoelastic model for VH [10]. $\lambda$, $\alpha$, and $\eta_p$ are the relaxation time, mobility factor, and polymer viscosity, respectively. The solvent viscosity is $\eta_s = 0.001$ Pa.s. The Deborah number of each mode, $De_n$, is calculated by Eq. (26).

| Mode ($n$) | $\lambda_n$ (s) | $\alpha_n$ | $\eta_{p,n}/\eta_s$ | $De_n$ |
|---|---|---|---|---|
| 1 | 0.01576 | 0.10954 | 140.09 | 18.70 |
| 2 | 3.00229 | 0.56796 | 18519.59 | 3562.28 |
| 3 | 0.10996 | 0.74892 | 324.60 | 130.47 |

The parameters of interest are the final droplet shape, the travel time ($\tau_{\text{tra}}$), which is defined as the time taken for the droplet to touch the retina surface, settling time ($\tau_{\text{set}}$), which is the next time period until the drop reaches its final shape over the retina surface, coverage area ($A_c$), which is defined as the final contact area between drop and retina, droplet surface area ($A_d$), and the distribution of the impact compressive stress ($p_{\text{FDT}} = -\sigma_{nn} = p - \tau_{nn}$), which is the compressive normal stress exerted on the retina surface by the FDT process.

### 2.2. Governing equations

Adopting a single-fluid formulation [31], the continuity and momentum equations for an incompressible immiscible two-phase flow without any mass transfer at the interface within the eye are:

$$\nabla \cdot \boldsymbol{u} = 0, \tag{1}$$

$$\frac{\partial (\rho \boldsymbol{u})}{\partial t} + \nabla \cdot (\rho \boldsymbol{u}\boldsymbol{u}) = -\nabla p + \nabla \cdot \boldsymbol{\tau} + \rho \boldsymbol{g} + \nabla \cdot \boldsymbol{\tau}_m + \boldsymbol{F}_\sigma. \tag{2}$$

Here, $\rho$, $\boldsymbol{u}$, $p$, $\boldsymbol{g}$, $\boldsymbol{F}_\sigma$, $\boldsymbol{\tau}$, and $\boldsymbol{\tau}_m$ are the mixture density, velocity, pressure, gravitational acceleration, surface tension per unit volume, viscous stress tensor, and magnetic stress tensor, respectively. In the present text, the bold symbols indicate tensorial quantities. The mixture properties are determined by

$$\rho = I_d \rho_d + (1 - I_d)\rho_c, \tag{3}$$

$$\boldsymbol{\tau} = I_d \boldsymbol{\tau}_d + (1 - I_d)\boldsymbol{\tau}_c, \tag{4}$$

where $I_d$ is the (droplet) phase-indicator function. The subscripts $d$ and $c$ refer to the droplet and carrier (VH) phase, respectively. The viscous stress tensor constitutive law for each phase is given in section 2.2.1, the magnetic force per unit volume ($\boldsymbol{F}_m = \boldsymbol{\nabla}.\boldsymbol{\tau}_m$) is introduced in section 2.2.2, and the front-tracking method for the calculation of $I_d$ and $\boldsymbol{F}_\sigma$ is described in section 2.2.3.

*2.2.1. The viscous stress tensor model*

The viscous stress tensor for the droplet Newtonian fluid phase ($\boldsymbol{\tau}_d$) is described by the Stokes law:

$$\boldsymbol{\tau}_d = \boldsymbol{\tau}_s \equiv 2\eta_s \boldsymbol{D}, \tag{5}$$

where $\eta_s$ is the fluid or solvent viscosity and $\boldsymbol{D}$ is the strain-rate tensor,

$$\boldsymbol{D} = \frac{1}{2}[\nabla \boldsymbol{u} + (\nabla \boldsymbol{u})^T]. \tag{6}$$

For the VH viscoelastic phase, the extra-stress tensor ($\boldsymbol{\tau}_c$) is decomposed into a solvent contribution, which is governed by Eqs. (5) and (6), and a polymeric contribution as:

$$\boldsymbol{\tau}_c = \boldsymbol{\tau}_s + \boldsymbol{\tau}_p. \tag{7}$$

For the polymeric stress tensor, $\boldsymbol{\tau}_p$, the m-mode Giesekus constitutive law [32] is:

$$\boldsymbol{\tau}_p = \sum_{n=1}^{m} \boldsymbol{\tau}_{p,n}, \tag{8}$$

$$\boldsymbol{\tau}_{p,n} + \lambda_n \check{\boldsymbol{\tau}}_{p,n} + \alpha_n \frac{\lambda_n}{\eta_{p,n}}(\boldsymbol{\tau}_{p,n}.\boldsymbol{\tau}_{p,n}) = 2\eta_{p,n}\boldsymbol{D}, \tag{9}$$

where $n$ is the viscoelastic mode index, $m$ the number of modes, $\lambda_n$ the relaxation time, $\alpha_n$ the mobility factor, $\eta_{p,n}$ the polymeric viscosity coefficient, and $\check{\boldsymbol{\tau}}_{p,n}$ the upper-convected time derivative of the polymeric stress tensor defined by

$$\check{\boldsymbol{\tau}}_{p,n} = \frac{\partial \boldsymbol{\tau}_{p,n}}{\partial t} + \boldsymbol{u}.\nabla \boldsymbol{\tau}_{p,n} - \boldsymbol{\tau}_{p,n}.\nabla \boldsymbol{u} - (\nabla \boldsymbol{u})^T.\boldsymbol{\tau}_{p,n}. \tag{10}$$

*2.2.2. The FerroHydroDynamics (FHD) model*

Assuming an incompressible linearly magnetizable medium, the magnetic stress tensor, $\boldsymbol{\tau}_m$ in Eq. (2), is given by [33-35]

$$\boldsymbol{\tau}_m = -\frac{1}{2}\mu|\boldsymbol{H}|^2\boldsymbol{I} + \mu\boldsymbol{H}\boldsymbol{H}; \ \mu = \mu_0(1+\chi), \tag{11}$$

where $\mu_0$, $\boldsymbol{H}$, and $\boldsymbol{I}$ are the magnetic permeability of vacuum, magnetic field vector, and identity tensor, respectively, $\chi$ is the mixture magnetic susceptibility, and the magnetic permeability of the mixture, $\mu$, is computed by

$$\mu = I_d\mu_d + (1 - I_d)\mu_c. \tag{12}$$

The operator $|.|$ returns the magnitude of a vector.

Assuming magnetostatics, $\nabla \times \boldsymbol{H} = 0$, the magnetic field can be expressed in terms of a magnetic potential, $\varphi$, as

$$\boldsymbol{H} = \nabla\varphi. \tag{13}$$

For non-magnetizable media and a multiphase flow of non-conducting linear material, $\nabla.\boldsymbol{B} = 0$, where $\boldsymbol{B} = \mu\boldsymbol{H}$ is the magnetic induction vector, and the magnetic potential is governed by

$$\nabla.(\mu\nabla\varphi) = 0. \tag{14}$$

In the present work, the magnets (magnetizable material) are not included in the computational domain and are accounted for by proper boundary conditions. Therefore, Eq. (14) is solved within a large cubic box which is composed of two regions: the eye and its surrounding environment, excluding the magnets volumes. These regions are coupled by the continuity of the magnetic potential and normal magnetic field at their interface [36]:

$$\varphi_1 = \varphi_2, \tag{15}$$

$$\mu_1\boldsymbol{n}_1.\nabla\varphi_1 = -\mu_2\boldsymbol{n}_2.\nabla\varphi_2, \tag{16}$$

where indices 1 and 2 indicate the regions on the two sides of the interface and $\boldsymbol{n}$ is the interface normal pointing into each region.

Using the current assumptions, it can be shown that the magnetic force per unit volume can be simplified to

$$\boldsymbol{F}_m = \nabla.\boldsymbol{\tau}_m = -\frac{1}{2}|\boldsymbol{H}|^2\nabla\mu. \tag{17}$$

### 2.2.3. The Front-Tracking Method (FTM)

For the calculation of the phase-indicator function, $I_d$, and surface tension, $\boldsymbol{F}_\sigma$, the FTM approach [31, 37] which uses a Lagrangian surface mesh, i.e., the front, is used. To track the location of each point at the interface between two phases, $\boldsymbol{x}_f$, we integrate

$$\frac{d\boldsymbol{x}_f}{dt} = \boldsymbol{u}_f = \boldsymbol{u}(\boldsymbol{x} = \boldsymbol{x}_f). \tag{18}$$

Then, the gradient of the indicator function, $G_I$, and $F_\sigma$ can be computed through the following transformations from the front to Eulerian fields:

$$G_I(x) = \int_{A_f} \delta(x - x_f) n_f dA_f, \tag{19}$$

$$F_\sigma(x) = \int_{A_f} \delta(x - x_f) f_{\sigma,f} dA_f, \tag{20}$$

where $\delta(x)$ is a 3D delta function, $n_f$ the front unit normal vector pointing outside the drop phase, $f_{\sigma,f}$ the surface tension per unit front area, and the integrations are performed over the front surface, $A_f$. The details of the numerical procedure to compute Eqs. (18)-(20) comprise the FTM strategy and are described in section 3. Finally, the indicator function field is constructed by solving a Poisson's equation:

$$\nabla^2 I_d = \nabla \cdot G_I. \tag{21}$$

*2.3. Dimensionless parameters*

The important independent dimensionless parameters, in the absence of an initially imposed drop velocity and negligible gravity effect, include the Laplace number (the ratio of surface tension to the viscous force) [5, 33],

$$\text{La} = \frac{\sigma \rho_d D_d}{\eta_d^2}, \tag{22}$$

the ratio of the magnetic to viscous forces, the magnetic Laplace number [5, 33],

$$\text{La}_m = \frac{\mu_d H_0^2 \rho_d D_d^2}{\eta_d^2}, \tag{23}$$

where $H_0$ is the characteristic magnetic field value chosen as $H_0 = \Delta\phi_0/L_{\text{mag}}$ here, and the ratio of fluid phase properties,

$$\frac{\rho_c}{\rho_d}, \frac{\eta_{0,c}}{\eta_d}, \frac{\mu_d}{\mu_c}, \tag{24}$$

where $\eta_{0,c}$ is the total (zero-shear-rate) viscosity of VH defined by

$$\eta_0 = \eta_s + \sum_{n=1}^{m} \eta_{p,n}. \tag{25}$$

Other dimensionless parameters related to the viscoelastic VH material are the Deborah number, the polymeric-to-solvent viscosity ratio, and the mobility factor defined for each mode,

$$\mathrm{De}_n = \frac{\lambda_n}{\tau_v}, \frac{\eta_{p,n}}{\eta_s}, \alpha_n, \tag{26}$$

where $\tau_v = \rho_c D_d^2/\eta_{0,c}$ is the viscous time scale of VH.

The most important geometrical dimensionless parameters are:

$$\frac{D_d}{D_{\mathrm{eye}}}, \frac{l_x}{D_d}, \frac{l_y}{D_d}, \frac{L_{\mathrm{mag}}}{D_d}. \tag{27}$$

Other dependent parameters can be related to the abovementioned ones, e.g., the ratio of the magnetic force to surface tension or magnetic Bond number,

$$\mathrm{Bo}_m = \frac{\mathrm{La}_m}{\mathrm{La}}. \tag{28}$$

The objectives introduced in section 2.1 can also be presented in dimensionless forms as:

$$\tau_{\mathrm{tra}}^* = \frac{\tau_{\mathrm{tra}}}{\tau_v}, \tau_{\mathrm{set}}^* = \frac{\tau_{\mathrm{set}}}{\tau_v}, A_c^* = \frac{A_c}{\pi D_d^2/4}, \zeta = \frac{\pi D_d^2}{A_d}, p_{\mathrm{FDT}}^* = \frac{p_{\mathrm{FDT}}}{\frac{1}{2}\mu_c H_0^2}, \tag{29}$$

where $\zeta$ is the droplet sphericity parameter. The other dimensionless variables are also given by $\boldsymbol{u}^* = \rho_c D_d \boldsymbol{u}/\eta_{0,c}$, $\phi^* = \phi/\Delta\phi_0$, $\boldsymbol{H}^* = \boldsymbol{H}/H_0$, and $\tau^* = t/\tau_v$.

## 3. Numerical method

For the present simulations, the "cfdmfFTFoam" FTM solver, developed in the CFDMF group [38] based on the OpenFOAM (OF) Foundation finite volume CFD package (www.openfoam.org) version 2.3.0, is utilized. For the triangulated front Lagrangian surface mesh, while following the general strategy of the FTM solver of PARIS simulator [39], many new front processing algorithms have been implemented to adapt the solver to general unstructured Eulerian grids. For the solution of the viscoelastic equations, the rheoTool library [40, 41] (https://github.com/fppimenta/rheoTool) is utilized and made compatible with OF 2.3.0. Following is a summary of the numerical methods used in the present work.

### 3.1. The numerical algorithms

The FTM is applied by six consecutive stages at each time step: 1) Front remeshing (coarsening, refining, and smoothing), 2) Surface tension computation, 3) Front-to-field communications, 4) Indicator function construction, 5) Front advection, and 6) Volume correction. The refining and coarsening algorithms control the size and skewness of the front triangle elements [31, 39]. By default, the criteria for element refining and coarsening are chosen to keep the ratio of the front element edge size to the Eulerian grid length scale within the range (0.25, 0.7) and a maximum allowable front element skewness of 1.5. The smoothing or undulation removal stage prevents front distortion due to numerical-error-induced noise and wiggles and plays a key role in the FTM robust performance. For the smoothing, three different methods, including 3D Trapezoidal Sub-grid Undulations Removal (TSUR3D) [42] and Volume Conserving Smoothing (VCS) III and IV [43], are implemented and compared. Note that the smoothing algorithm is executed every $N_s$ time steps, where $N_s$ should be chosen carefully for each problem to simultaneously maintain the robustness and accuracy of the whole algorithm. The influence of those parameters for the present problem is examined in sections 4.1.1 and 4.1.2. In step 2, the net surface tension at each triangle element, $\boldsymbol{f}_{\sigma,f} dA_f$ in Eq. (20), is computed by the direct element-based algorithm [44, 45].

In step 3, the surface tension and the gradient of the indictor function are transformed from the front elements to the Eulerian grid by Eqs. (19) and (20). To implement this stage for a general unstructured Eulerian grid, the 1st-order Reproducing Kernel Particle Method (RKPM), which has been used previously in the immersed boundary method [46-48] and for the coarse-graining in Eulerian-Lagrangian simulations [49], is adopted. The advantage of RKPM compared to common methods already used in FTMs is the preservation of the torque (the first-order moments) in addition to the force during the distribution of surface tension to the Eulerian grid. In our RKPM, the front data is transformed from a front vertex to the Eulerian grid cells in the region of influence of that vertex. The region of influence is taken to be a sphere with radius $\alpha h_k$ where $\alpha = 2$ by default [48] and $h_k$ is the characteristic length scale at the k$^{th}$ front element. Here, $h_k$ is set to the averaged cell size (the cubic root of cell volume) over all Eulerian grid cells in the bounding cubic box of the droplet at each instance of time. For an efficient implementation of RKPM and its search algorithms within the region of influence, readers are referred to the algorithms presented in our previous work [49].

At step 4, the indicator function is reconstructed by solving Eq. (21) using the second-order "Gauss linear" discretization scheme [50] and proper boundary conditions at all boundaries. A zero-valued boundary condition, i.e., $I_d = 0$, works for all problems considered in the present study since we keep a thin layer of carrier fluid between the wall and the droplet surface. For the main FDT-RD problem, this means that the retina is assumed to be a hydrophobic surface for the droplet material. For step 5, the locations of front points are updated by the available OF face-to-face procedure for Lagrangian particle tacking in general unstructured grids [51] using a Lagrangian Courant number of 0.25. The fluid velocity on the right-hand-side of Eq. (18) is interpolated at the front point, $x_f$, using the "cellPoint" algorithm, where, first, the fluid velocities are computed on the Eulerian grid vertices using the linear interpolation scheme, then, the fluid velocity at the front point location within an Eulerian grid cell is estimated by cell decomposition into tetrahedrons and using an interpolation based on the Barycentric coordinates [52]. The details of the algorithm was provided in the supplementary material of our previous publication [49]. The explicit Euler scheme is used for the integration of Eq. (18), which is reasonable owing to the small time steps used in the simulations.

In step 6, an ad hoc algorithm is used to compensate for the inherent shortcoming of the Lagrangian models (like FTM) in conserving volume/mass. The details of the algorithm used in this study are presented in Appendix A, and its importance for the present application is highlighted in section 4.1.2.

To prevent HWNP, the conventional BSD approach [41] was tested and found insufficient in our main FDT-RD case. Therefore, for the robust solution of the viscoelastic equations, the log-conformation tensor approach [53] and the stress-velocity coupling method described in detail in reference [40] are employed. The pressure-velocity coupling is treated using the Pressure-Implicit with Splitting of Operators (PISO) algorithm [54] with 3 pressure corrector loops. The second-order "Gauss linear" [50] discretization is adopted for the momentum advection and all diffusion terms and the "Gauss linear" scheme for gradient operators, while the polymeric stress advection term is discretized by the "GaussDefCmpw cubista" scheme [40, 41]. The implicit first-order Euler scheme is used for the time derivatives while a maximum Courant number of 0.25 is maintained for the dynamic calculation of time steps during the simulations. The systems of discretized equations are solved by the Geometric Agglomerated Algebraic Multi-Grid (GAMG) solver with the GaussSeidel smoother for the pressure and the Stabilized Preconditioned Bi-Conjugate Gradient (PBiCGStab) algorithm with the Diagonal Incomplete LU decomposition (DILU) pre-conditioning for the velocity and polymeric stress. The solution convergence at each time step is ensured by setting the residual tolerance of momentum, pressure, magnetic potential, and polymeric stress to $10^{-8}$, $10^{-8}$, $10^{-10}$, and $10^{-8}$, respectively.

The multi-region computational domain and the Eulerian grid are shown in figure 2. The multiphase flow equations are solved in the eye domain with a no-slip velocity and zero-gradient pressure at the eye surface (indicated in green in figure 2a). The magnetic field equations are solved within both the eye domain and its surrounding medium, which is bounded by a sphere of diameter $3D_{eye}$. A zero-gradient magnetic potential is imposed at the far-field boundary (indicated in blue in figure 2a), a positive fixed-value magnetic potential of $\Delta\phi_0/2$ is set at the upper face of each magnet and a negative fixed-value magnetic potential of $-\Delta\phi_0/2$ at their lower faces. The side faces of magnets are magnetically insulated (a zero-gradient potential condition). The coupling between the magnetic field inside and outside of the eye is treated using Eqs. (15) and (16) at the eye surface. In each time loop, the magnetic field equations, Eqs. (12)-(14), in each region along with the inter-region boundary conditions, Eqs. (15) and (16), are solved in an iterative fashion. About four iterations are necessary to achieve convergence of the magnetic field across all regions. Then, the magnetic force density, Eq. (17), is updated and used in the FTM equations. An O-type block-structured grid of hexahedral cells and a finer resolution near the walls is used within the eye domain. The grid outside the eye is also mostly structured except in a small region near the magnets and lower part of the surrounding domain, which is filled with unstructured tetrahedral cells with the finest resolution near the magnets.

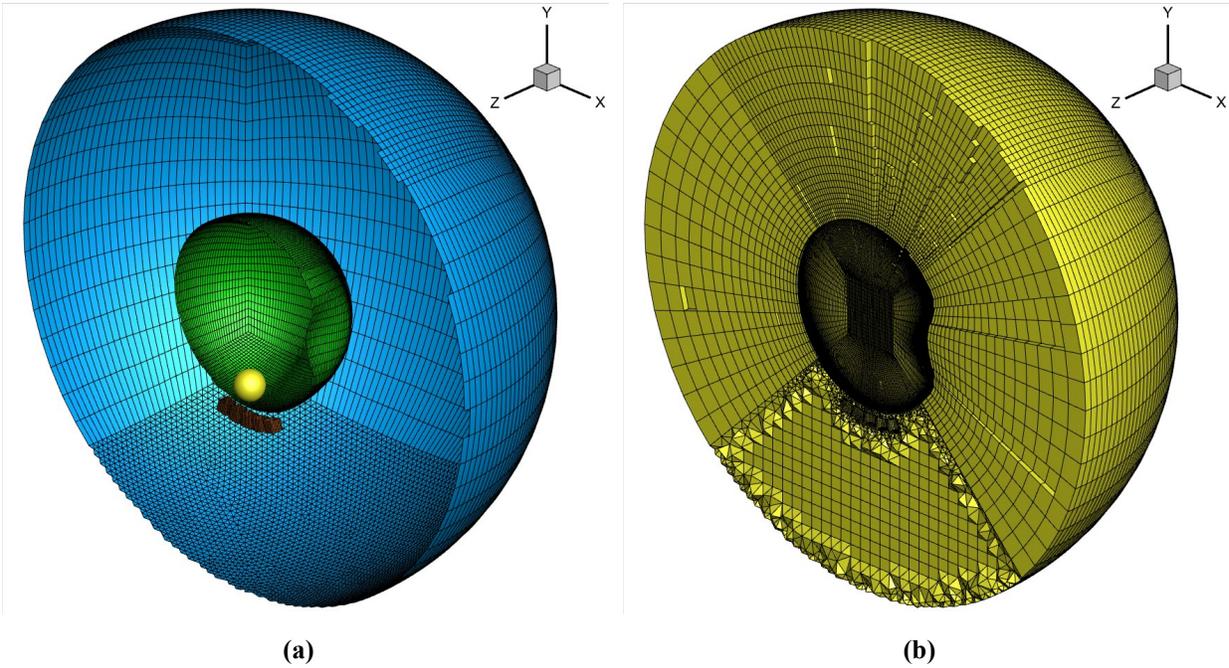

(a) (b)
**Figure 2** A half of the multi-region grid: a) The surface of the magnets, drop, eye, and surrounding medium are indicated in red, yellow, green, and blue, respectively. The drop and magnets were left uncut for clarity. b) The volumetric grid topology.

## 4. Results and discussion

To validate the FTM framework for FDT in a real eye geometry and to address the challenges of extending the FTM to complex geometries using a general 3D unstructured grid, several benchmarks and tests are described in section 4.1. Then, the best setup is used to study the FDT for RD treatment in section 4.2.

### 4.1. The solver evaluation: validation and numerical considerations

For the validation of the two-phase flow solver and the interface tracking strategy in the absence of magnetic field and viscoelastic effects, droplet migration and deformation in creeping flow within a circular tube is considered in benchmark 1 (section 4.1.1). In addition, it is shown that the smoothing step is critical and a comparison between the performance of different smoothing algorithms is made to choose the best approach. In benchmark 2 (Appendix B), the FHD submodel is validated by the prediction of the deformation of a ferrofluid droplet in a uniform external magnetic field. Then, in benchmark 3 (Appendix C), the model for the rheological behavior of VH is assessed by predicting a viscoelastic fluid flow with the same constitutive law through a contraction. And in benchmark 4 (section 4.1.2), additional numerical challenges imposed by the motion of a droplet towards a wall are addressed.

#### 4.1.1. The droplet deformation in creeping flow within a circular capillary tube

Figure 3 shows an initially spherical droplet of diameter $D_d$ moving along the axis of a circular capillary tube of diameter, $D_t$, due to the imposed flow of a carrier fluid with a bulk velocity of $u_b$. Both droplet and carrier phase are Newtonian fluids. The dimensionless parameters governing the fully-developed shape of the droplet are:

$$\text{Re} = \frac{\rho_c u_b D_t}{\eta_c}, \text{Ca} = \frac{\eta_c u_b}{\sigma}, \frac{D_d}{D_t}, \frac{\rho_c}{\rho_d}, \frac{\eta_c}{\eta_d}, \tag{24}$$

where Re and Ca are the Reynolds and capillary numbers, respectively. Under the creeping flow condition and for a small droplet far from the pipe walls, i.e., small $D_d/D_t$ ratios, Re and $D_d/D_t$ are irrelevant and an analytical solution has been found for the fully-developed shape of the droplet by Nadim and Stone [55]. For larger droplets experiencing limited wall effects, experimental measurements have been conducted by Olbricht and Kung [56]. Both conditions are considered here with the parameters given in table 2.

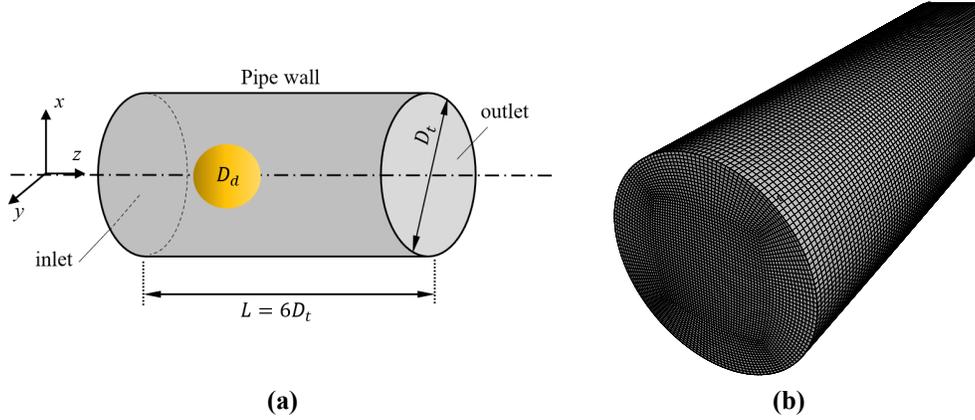

(a)                              (b)

**Figure 3** Benchmark 1: (a) The schematic geometry and computational domain, and (b) The block-structured Eulerian grid.

Table 2 Benchmark 1: The non-dimensional parameters.

| Case | Re | Ca | $\eta_c/\eta_d$ | $\rho_c/\rho_d$ | $D_d/D_t$ | Reference |
|---|---|---|---|---|---|---|
| Case 1 | 0.1 | 1.0 | 1.0 | 1.0 | 0.3 | [55] |
| Case 2 | 0.1 | 0.1 | 1.0 | 1.0 | 0.9 | [56] |

The 3D computational domain used for benchmark 1 is shown in figure 3a. A block-structured O-type grid, shown in figure 3b, with 1489600 grid cells for case 1 and 748800 cells for case 2 is used, i.e., 20 and 48 Cells Per initial droplet Diameter (CPD) for case 1 or 2, respectively. A uniform velocity profile is assumed at the inlet which quickly becomes fully-developed due to the small Reynolds number of the flow. The no-slip condition at the walls and fixed pressure at the outlet are applied. The simulations are continued until the droplet reaches its final unchanged shape.

In most of the cases studied in the present work, it is observed that the smoothing process is a key factor in the robustness of the solution. In the present benchmark, case 2 is more challenging since the drop interface moves near the pipe walls and the effect of the wall and the near wall high-velocity-gradient region on the front is significant. When no smoothing is applied, figure 4a, a considerable level of grid-scale undulations is observed on the rear surface of the droplet. This growing noise on the front is a feature of front tracking algorithms and stems mostly from the Lagrangian advection of the front points. To remove these undulations, we tested three smoothing or undulation removal algorithms, i.e., TSUR3D, VCS III, and VCS IV, for the current test case. The smoothing interval was chosen as $N_s = 50$ for all algorithms.

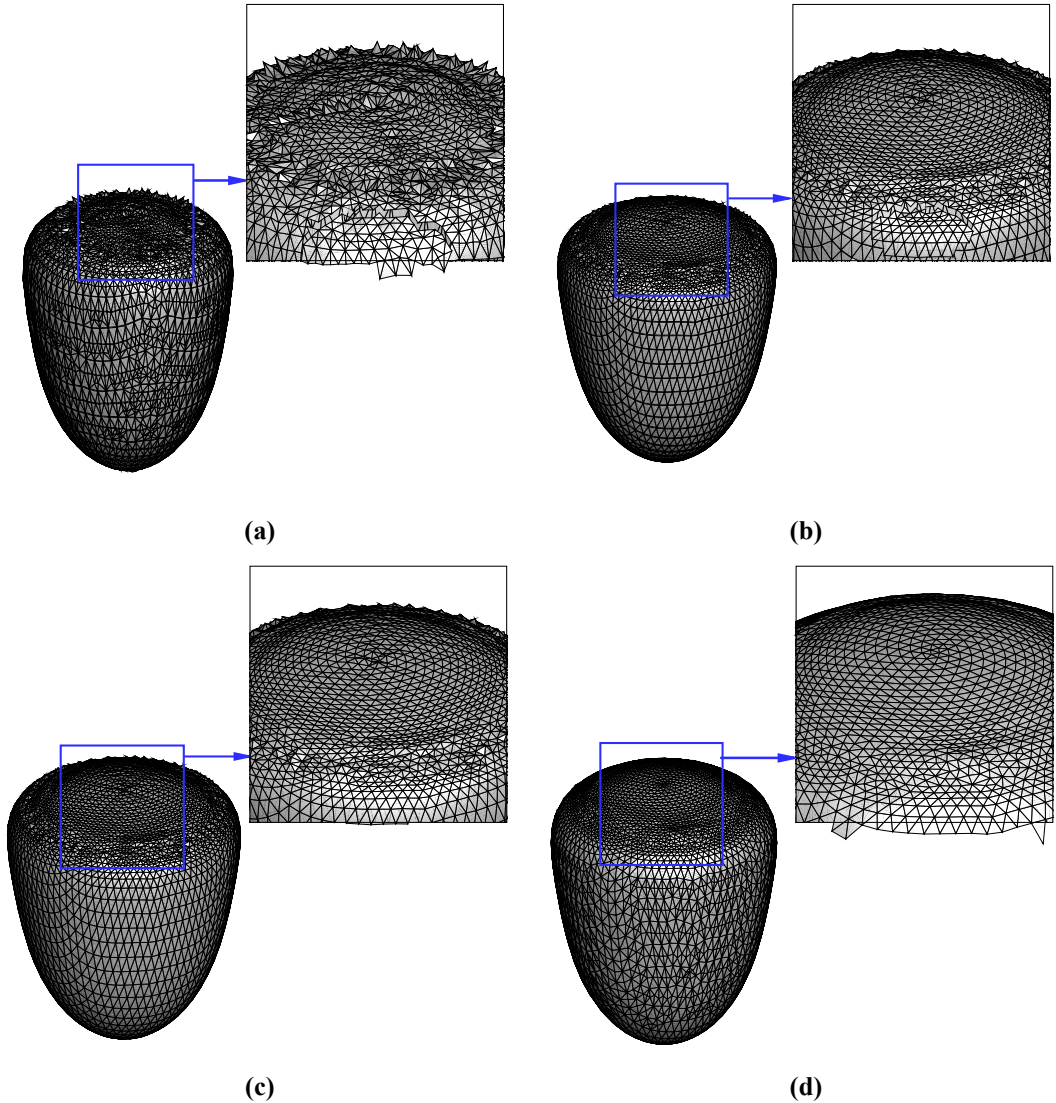

**Figure 4** Benchmark 1: The fully-developed topology of drop for case 2 using different smoothing algorithms. a) No smoothing, b) TSUR3D, c) VCS III, and d) VCS IV.

As can be observed in figure 4b and c, the results of using TSUR3D and VCS III are very similar to each other. Although these methods noticeably smoothed the jagged areas, they did not successfully attenuate front undulations in this case (and many other cases of the present work). Note that the solutions with these two smoothing methods as well as the one with no smoothing diverged when we continued the run for a long time. On the other hand, incorporating VCS IV, figure 4d, resulted in a robust solution with effective attenuation of the grid-scale noises. For a more quantitative assessment, the dimensionless droplet surface area variation over time for different smoothing algorithms is reported in figure 5. According to this figure, using VCS IV, droplet surface area approaches a constant fully-developed value after about 5 seconds. On the contrary, the incorporation of the other methods leads to the continual growth of the drop surface area even after 5 seconds due to the increasing undulations on the droplet surface, which eventually ends in solution divergence for all of these cases.

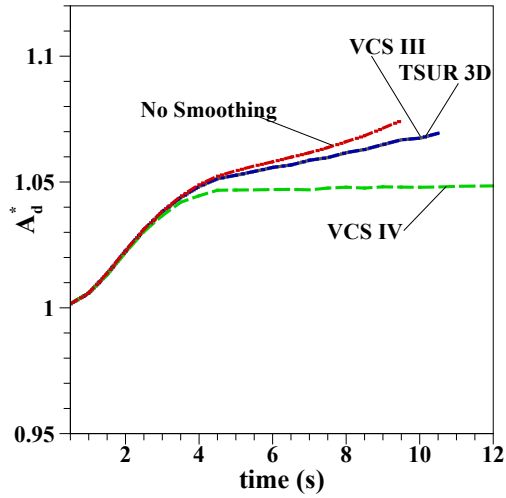

**Figure 5** Benchmark 1: The dimensionless droplet surface area ($A_d^* = A_d/(\pi D_d^2)$) over time for case 2 using different smoothing algorithms.

To justify these results, the algorithms of these methods should be compared. All algorithms are volume conserving. In TSUR3D and VCS III, an explicit Laplace smoothing on each front vertex is applied which is probably the cause of their similar performances. While VCS IV incorporates a simultaneous Laplace smoothing for both end-vertices of a front edge. More importantly, VCS IV imposes an additional constraint of minimizing the front edge translation in the front-normal direction which is deemed to be the main cause of smoothing the pointed protrusions observed with other algorithms. We made the same conclusion of the superiority of VCS IV algorithm in other benchmarks and this method is used by default for the rest of the results reported in this paper, unless stated otherwise. It is worth noting that the smoothing interval is an important parameter too and over-smoothing the front can reduce the solution accuracy by attenuating the physical high-curvature regions, especially for VCS IV which possesses a stronger smoothing character. To consider this fact, the choice of $N_s$ should be made carefully for each case in such a manner that the smoothing-independent results are obtained while the grid-scale undulations are suppressed. This is discussed further in section 4.1.2.

Figure 6 shows a comparison of the fully-developed topology of the droplets of case 1 and 2 using the present FTM and the analytical and experimental results. According to this figure, our FTM simulations are in reasonable agreement with the reference results, supporting the validity of our FTM solver.

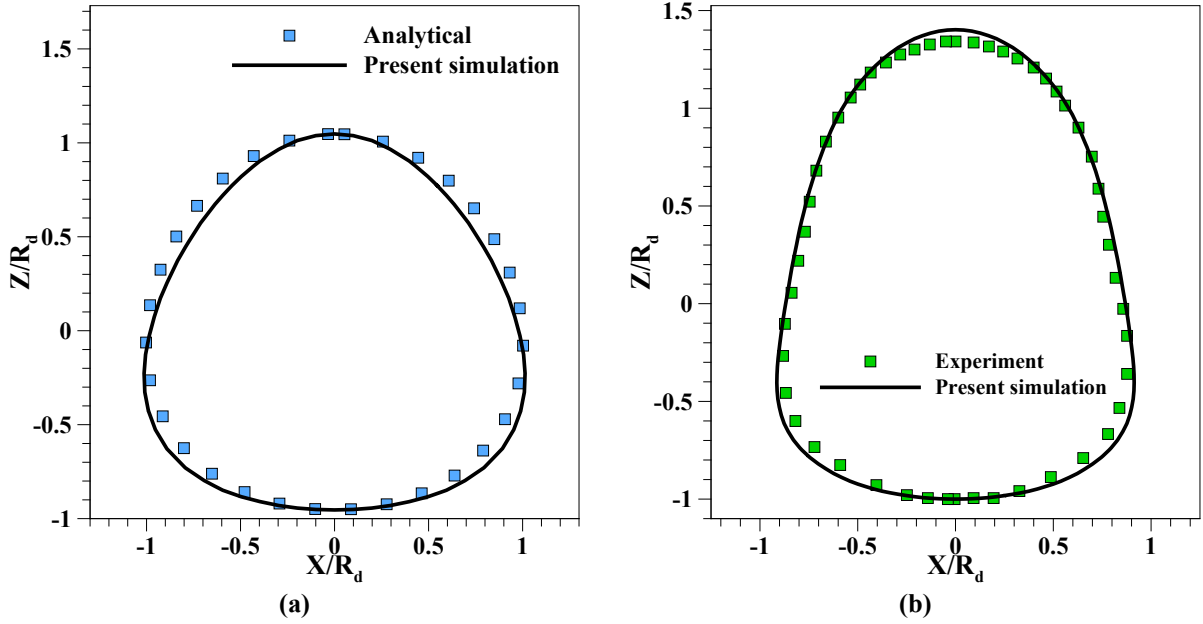

**Figure 6** Benchmark 1: The fully-developed topology of drop for a) case 1 and b) case 2. Comparison of the of the present FTM solutions against the analytical solution [55] and experimental measurements [56].

*4.1.2. Numerical considerations for the use of FTM for a real eye problem*

To use the FTM solver for FDT in the real eye geometry, several numerical considerations and challenges remain, which are addressed in this section. These include 1) the design of the Eulerian grid and grid-independence test, 2) the choice of a threshold distance, 3) the appropriate frequency of the smoothing step or undulation removal, and 4) the necessity and performance of the modified volume correction algorithm.

*4.1.2.1.The grid design, independence test, and threshold distance*

The grid topology is shown in figure 2. A non-uniform grid with a small expansion factor of 1.036 in the wall-normal direction is used for efficient computations. Among the seven blocks of the eye grid, the resolution of the grid is finer in the lowest block, where the drop resides and moves towards the eye wall (see also figure 7).

One of the major challenges in our study was the accuracy and robustness of the solver when the droplet gets very close to or is in contact with the walls. This can lead to divergence in the solution due to abrupt changes in the calculated magnetic force or the robustness of the FTM sub-algorithms. To address this issue, a tiny distance, called the threshold distance, $\varepsilon$, is defined (see figure 7). After step 5 of the FTM solver, the location of each front node that is closer to walls than the threshold distance is corrected in the wall-normal direction to keep the front nodes at least $\varepsilon$ away from walls. Additionally, the Eulerian grid cell(s) within the threshold distance, $\varepsilon$, from a wall are further refined to have 3 layers of cells between the threshold and the eye surface. The inclusion of this tiny distance considerably improves the robustness of the FTM steps; in addition, considering at least 3 layers of cells between the threshold and wall is crucial for the accuracy of the indictor function construction, using the method of Poisson's equation, and more importantly for the solution of the magnetic field and force, when the drop is in contact with walls.

It is important to check the independence of the solution from the value of the threshold distance, $\varepsilon$. Figure 8a shows a comparison of two objective parameters for different $\varepsilon/D_d$ values. The maximum difference between the predictions using $\varepsilon/D_d = 1/50$ and $1/100$ is less than 1%, therefore, $\varepsilon/D_d = 1/50$ is chosen for the rest of simulations.

Different grid resolutions were examined to find the optimal choice for the present problem. The specification for the three finest grid we tested along with the CPU time (on a 12-core Intel(R) Core(TM) i7-6800k CPU @ 3.40 GHz) for each simulation are given in table 3. In each case, the number of Eulerian grid Cells Per initial drop Diameter (CPD) is reported at the initial position of the drop. This grid resolution and CPD increases as the drop gets closer to the wall.

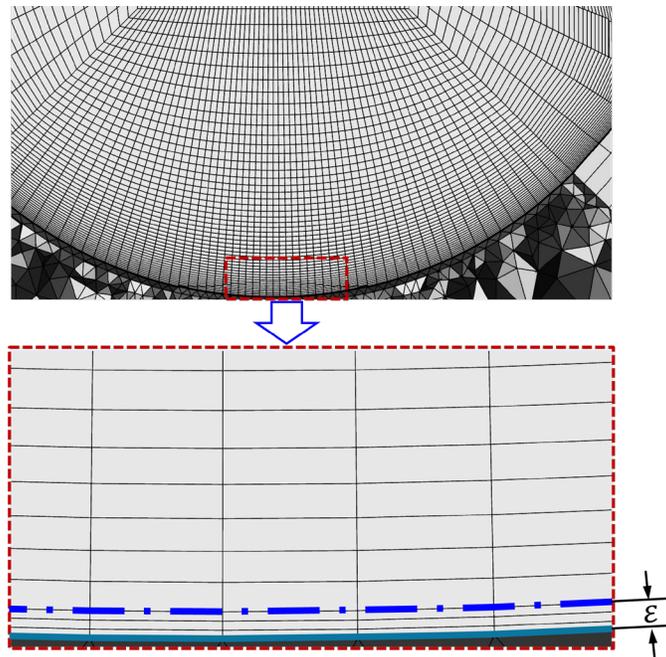

**Figure 7** A view of the Eulerian grid, showing the lowest block of the eye (up), and a zoomed-in region near the wall, displaying the threshold surface indicted by the blue dash-dotted line (down).

**Table 3** The grid specification and predicted travel time: CPD stands for "number of Cells Per initial droplet Diameter".

| Grid name | Cells | CPD | $\tau_{tra}$ (s) | Error (%) | CPU time (s) |
|---|---|---|---|---|---|
| Coarse Mesh | 450,000 | 19 | 1.942 | 1.46 | 69,718 |
| Medium Mesh | 900,000 | 24 | 1.967 | 0.32 | 152,729 |
| Fine Mesh | 1,700,000 | 30 | 1.971 | - | 290,301 |

The dependence of all objective parameters on the Eulerian grid resolution has been checked carefully. For instance, according to table 3, the relative error of the predicted travel time, $\tau_{tra}$, for the medium grid with CPD = 24 compared to the one with the finest mesh is smaller than 1%. Figure 8b shows the drop sphericity and (dimensionless) coverage area variation by time for the three grid resolutions. Again, the maximum deviation between the results with CPD = 24 and 30 is less than 1%. Additionally, the droplet shape and the fluid velocity contours at a vertical cross-section for different grid resolutions at two instances of time are compared in figure 9. At $t = 2\ s$, the shape of the droplet with CPD = 19 deviates from the shapes predicted with CPD = 24 and 30, while these two latter grid resolutions results are in very close agreement at all simulation times. Based on these tests, the medium grid resolution with CPD = 24 is selected for the rest of our analyses.

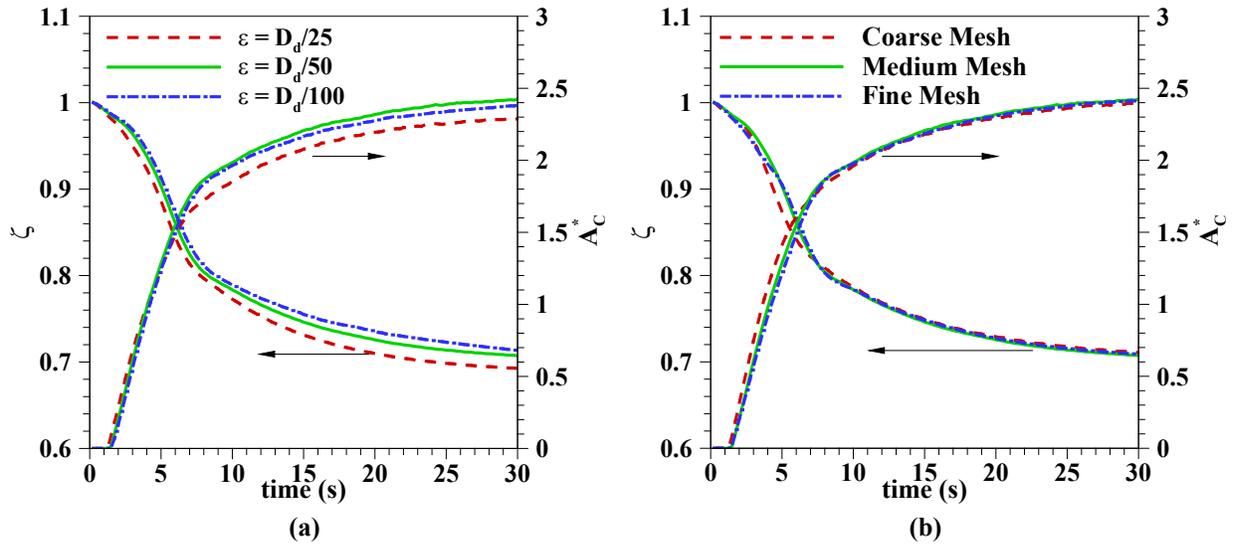

**Figure 8** The drop sphericity ($\zeta$) and (dimensionless) coverage area ($A_c^*$) over time for different a) (dimensionless) threshold distance values (medium grid) and b) grid resolutions ($\varepsilon/D_d = 1/50$).

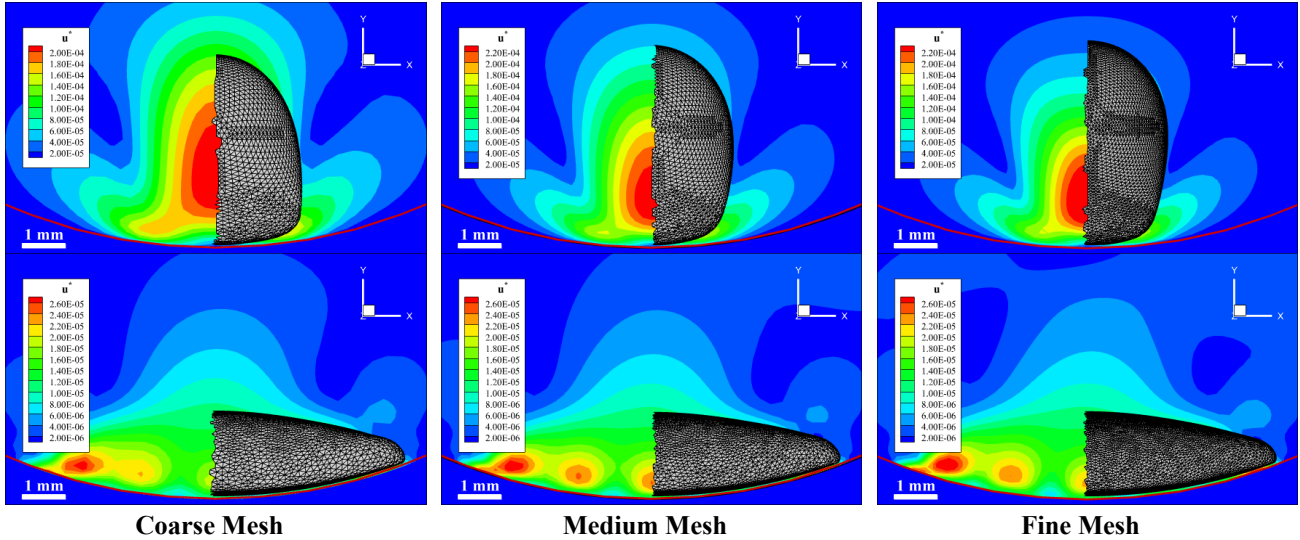

**Figure 9** The grid study: The (dimensionless) fluid velocity contours in $z = 0$ cross-section and (a half of) the droplet shape, at $t = 2$ s (top row) and 16 s (bottom row). The eye surface is indicated by the red solid line.

*4.1.2.2.The undulation removal interval*

In addition to the type of smoothing algorithm, which is selected as the VCS IV method for the main problem, the smoothing frequency or Undulation Removal Intervals (URI) is another important numerical parameter in FTM, which has to be chosen to maintain the solution stability and, at the same time, the solution independence from the particular choice of this parameter. For this purpose, URI = 10, 20, 40, and 60 are examined and the results are presented in figure 10. The result with URI = 10 is slightly different from the other ones, especially at the leading edge of the spreading drop on the eye surface. Due to the excessive surface smoothing, the leading edge of the drop predicted using URI = 10 has a smaller curvature compared to the other results. The results using URI = 20, 40, and 60 are in fine agreement with each other. However, to guarantee the solver robustness while maintaining its accuracy at high curvatures, the intermediate value of URI = 40 is chosen for the rest of simulations.

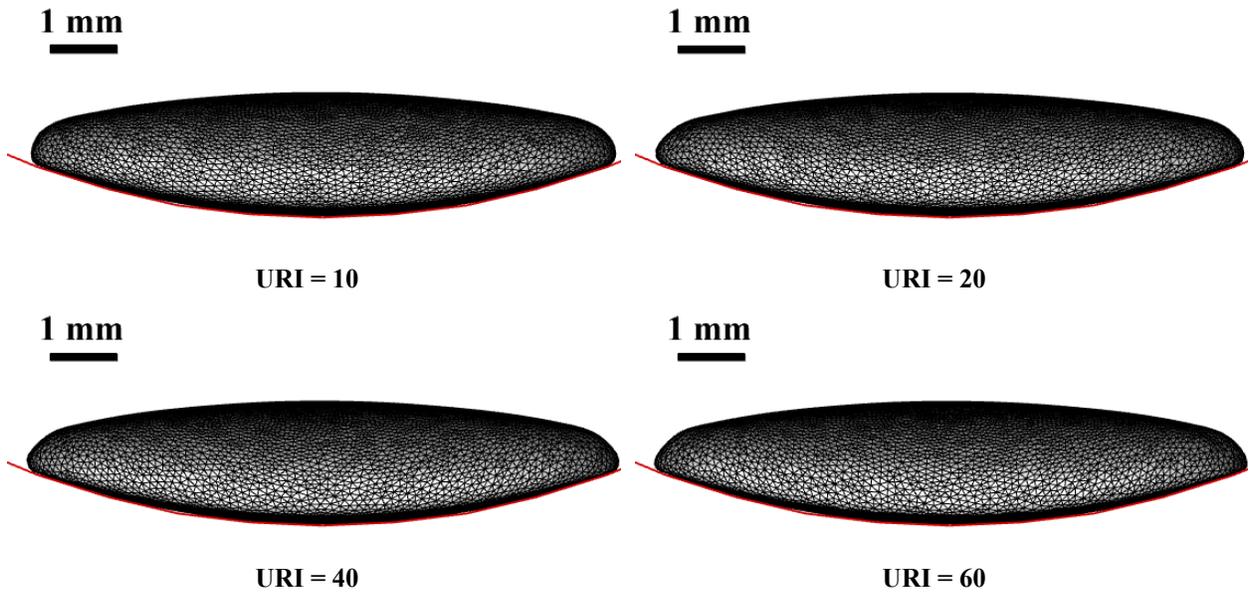

**Figure 10** The droplet (front mesh) shape at $t = 16\ s$ with URI = 10, 20, 40, and 60.

*4.1.2.3.The volume correction algorithm*

Due to the flow complexity and the presence of steep velocity gradients in near wall regions, the application of a velocity correction algorithm is critical for the current application. In this section, the performance of the volume correction algorithm, slightly modified to handle front mesh in a close distance to walls (see Appendix A), is assessed. Figure 11 compares the dimensionless drop volume versus time with and without the implementation of the volume correction. Without the volume correction, until about $t = 2\ s$ when the drop touches the eye surface, the volume conservation error is insignificant. After that, the drop volume is rapidly decreases due to advection errors, mainly induced by the inconsistencies between the velocity field on the Eulerian grid and its interpolated value at the front nodes. This leads to about 40% volume loss at the end of the simulation when the drop is settled on the eye surface, which is an unacceptable error. On the other hand, the incorporation of the volume change algorithm effectively removes this issue and is a must for the present application.

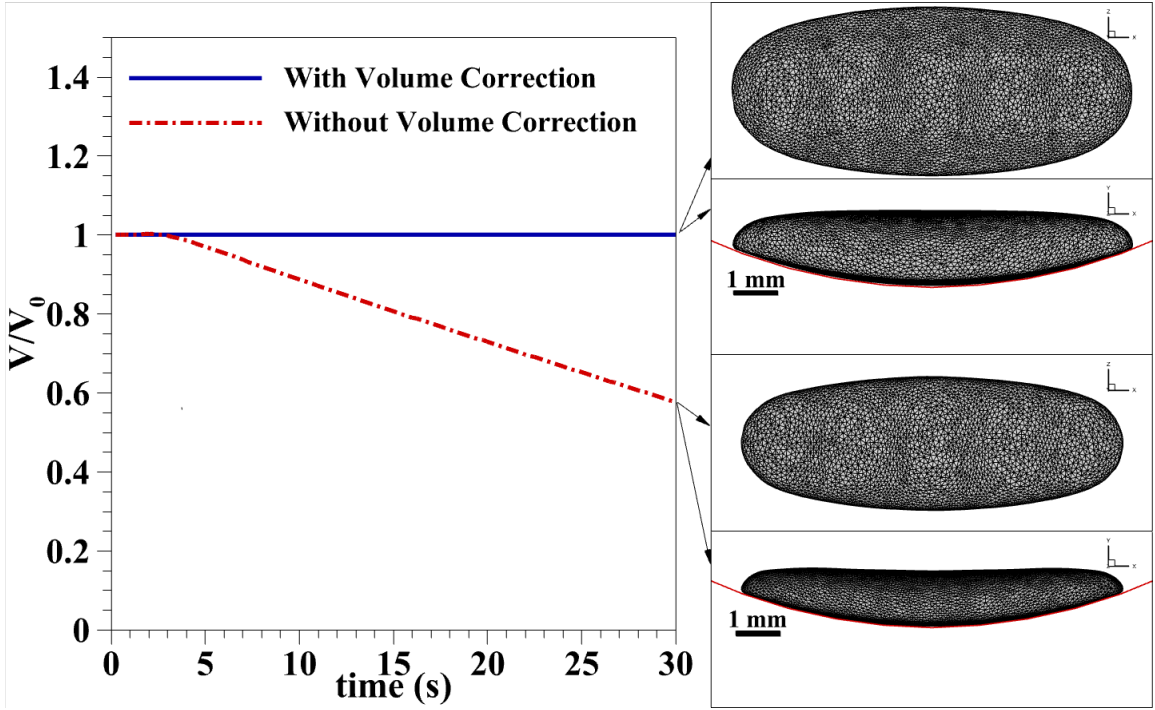

**Figure 11** The normalized drop volume over time with and without the volume correction algorithm. The right subfigures show the drop shapes at $t = 30\ s$, from a side and bottom view.

*4.2. The FDT for the treatment of RD in a real eye geometry*

Among the governing dimensionless parameters introduced in section 2.3, $La_m$ and La, which are the ratio of the magnetic to viscous and surface tension to viscous forces, respectively, are the main concern. It is anticipated that major dynamic objective parameters and characteristics of the system, like droplet dimensionless travel time, settling time, and shape change, significantly depend on $La_m$ and La. On the other hand, for stationary characteristics, such as (dimensionless) final retinal coverage area, final droplet shape, and final impact stress, the viscous force is irrelevant and these parameters are primarily functions of the magnetic Bond number, the ratio of the magnetic to interfacial forces. It should be noted that the elastic stress component of the VH, characterized by the Deborah number, can play a role even in stationary characteristics due to the memory effect, however, we considered a given VH with constant viscoelastic properties and Deborah number. To analyze the effect of $La_m$, La, and $Bo_m$ on the objective parameters, several cases are defined in table 4. Based on the problem statement in section 2.1, the values of the other dimensionless parameters are kept unchanged and equal to:

$$\frac{\rho_c}{\rho_d} = 0.76, \frac{\eta_{0,c}}{\eta_d} = 0.24, \frac{D_d}{D_{\text{eye}}} = 0.16, \frac{l_x}{D_d} = 0.0, \frac{l_y}{D_d} = 0.75, \frac{L_{\text{mag}}}{D_d} = 0.375.$$

The three rheological dimensionless parameters, $\alpha_n$, $\eta_{p,n}/\eta_s$, and $De_n$, corresponding to each viscoelastic mode of VH, were also given in table 1. In cases 1-4 in table 4, the Laplace number is changed by the variation of the surface tension coefficient, $\sigma$. In cases 1 and 5-7, the magnetic Laplace number is varied by changing the magnetic permeability of the droplet, $\mu_d$, in such a way that the magnetic Bond numbers of cases 5-7 are equal to those of cases 2-4, respectively.

**Table 4** The specification of cases for the parametric study of FDT-RD treatment.

| Case | $\mu_d/\mu_c$ | La ($\times 10^5$) | $La_m$ ($\times 10^2$) | $Bo_m$ ($\times 10^{-3}$) |
|---|---|---|---|---|
| 1 | 1.6 | 1.11 | 9.51 | 8.542 |
| 2 | 1.6 | 0.891 | 9.51 | 10.678 |
| 3 | 1.6 | 0.668 | 9.51 | 14.237 |
| 4 | 1.6 | 0.446 | 9.51 | 21.355 |
| 5 | 2.0 | 1.11 | 11.9 | 10.678 |
| 6 | 2.7 | 1.11 | 15.9 | 14.237 |
| 7 | 4.0 | 1.11 | 23.8 | 21.355 |

Figure 12 shows the variations of the travel time and settling time as functions of La, $La_m$, and $Bo_m$. The travel time is measured as the time period between the start of simulation and the instance when the first front node reaches the threshold surface. The settling time is calculated as the time span between the latter instance and the stationary state, which is chosen as the instance when the slop of the dimensionless droplet coverage area versus time falls below 2.5%. According to figure 12a, at constant $La_m$, the settling time continuously decreases with the increase in La while the travel time does not change considerably. This is due to the increase in the surface tension which reduces the drop shape change while the drop is settling on the retina. Based on figure 12b, by increasing $La_m$, at constant La, both the travel and settling times sharply decrease with a saturation-like behavior at large $La_m$ values. The decrease of the travel time by the rise in $La_m$ has also been reported for a drop in a Newtonian matrix and a simpler axisymmetric computational domain by Afkhami, et al. [5]. This is attributed to the larger propulsive magnetic force compared to the viscose friction. The saturation-like behavior of the settling time at large $La_m$, seen in figure 12b, is justified by analyzing the drop deformation metrics in figure 13b. The extent of the terminal deformation and the rate of deformation have opposing effects on the settling time. As $La_m$ and $\mu_d/\mu_c$ increase simultaneously, the drop experiences a larger terminal deformation while the deformation process gets a higher pace, see the increasing slopes of the deformation curves in figure 13b. On the other hand, when only La varies at constant $\mu_d/\mu_c$, the rate of shape change (the slope of the curves in figure 13a) does not vary much for the range of parameters in our study.

Important shape parameters are the dimensionless coverage area ($A_C^*$) and the drop sphericity ($\zeta$) which are plotted versus time in figure 13. The shape evolutions within the settling period for all cases show two distinct regimes: an initial fast shape change followed by a slow shape variation. The point of transition between the two regimes depends on $La_m$ (figure 13b) but does not show noticeable dependence upon La (figure 13a). The shape change depends on both La and $La_m$, however, at the terminal stationary state, the viscous force is irrelevant and the terminal dimensionless coverage area and sphericity are only functions of $Bo_m$. In figure 13a, by increasing $Bo_m$ from 8542 to 21355, the terminal coverage area grows from 2.42 to 3.66 (51% growth), while the same increase in $Bo_m$ in figure 13b results in a 19% growth of the coverage area. This is because in figure 13a (cases 1 and 5-7), in addition to $Bo_m$, $\mu_d/\mu_c$ increases; while in figure 13b (cases 1-4), it is unchanged. Therefore, $\mu_d/\mu_c$, has a significant effect on the terminal shape metrics and has to be considered along with $Bo_m$ to determine the final metrics. Additionally, $La_m$, rather than $Bo_m$, controls the shape change rate and spreading speed on the ocular surface, since in figure 13b, by increasing $Bo_m$ while keeping $La_m$ unchanged, no significant variation is observed in the slope of the curves. On the contrary, in figure 13a, the slope of the curves alters significantly as $La_m$ varies.

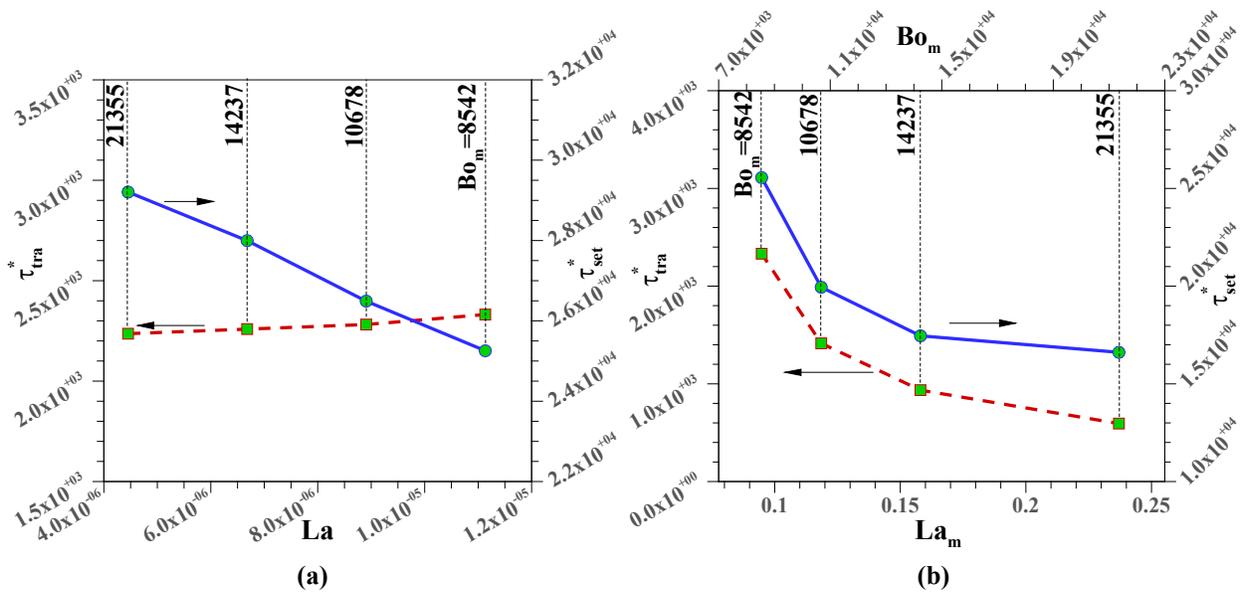

**Figure 12** The (dimensionless) travel and settling times versus the Laplace number (cases 1-4) (a) and magnetic Laplace number (cases 1, 5-7) (b).

According to figure 13b, as $La_m$ grows, a droplet recoil-like phenomenon occurs before the second stage of drop settling on the wall, e.g., see case 3 ($Bo_m = 14237$) around $\tau^* = 0.6 \times 10^4$ or case 4 ($Bo_m = 21355$) around $\tau^* = 0.4 \times 10^4$. This phenomenon is not observed in the other scenario (figure 13a), where $Bo_m$ increases with the decrease in La.

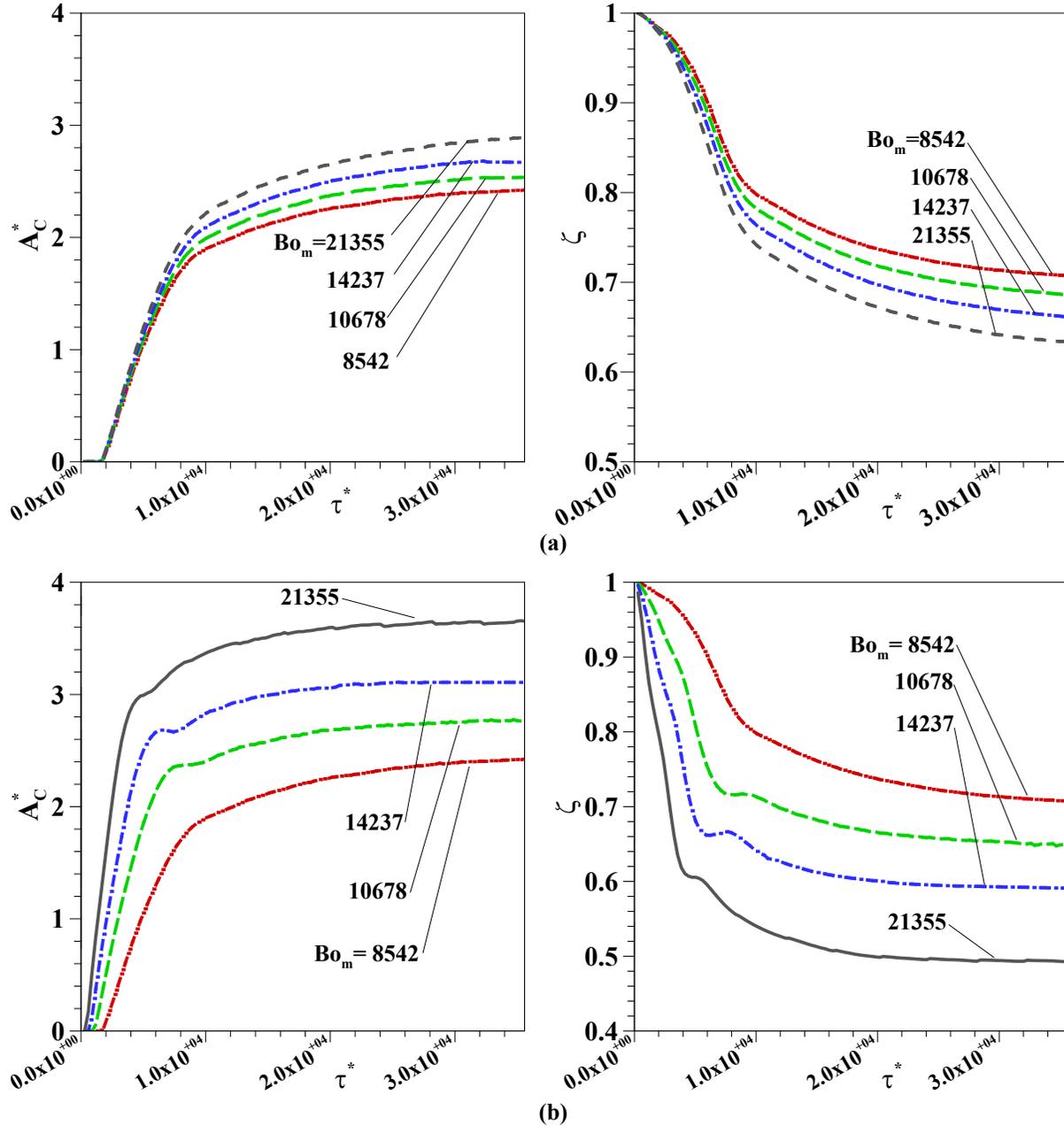

**Figure 13** The dimensionless retinal coverage area ($A_C^*$) and sphericity ($\zeta$) over time for different magnetic Bond numbers. (a) Only $Bo_m$ (or La) varies (cases 1-4), (b) $Bo_m$ (or $La_m$) and $\mu_d/\mu_c$ vary simultaneously (cases 1, 5-7).

To justify these findings, the droplet shape evolutions are compared in figure 14 and figure 15 from two views. By an increase in $La_m$, the dop impact velocity increases since the travel time considerably reduces, while this is not the case with the corresponding decrease in La, see $\tau^* = \tau_{tra}^*$ in figure 14. In addition, by increasing $Bo_m$ and $\mu_d/\mu_c$ simultaneously, a deep crease appears on the top surface of the spreading drop, which also persists in the final shape of the droplet, see case 7 in figure 15. This crease is not seen when only $Bo_m$ increases (case 4 in figure 15).

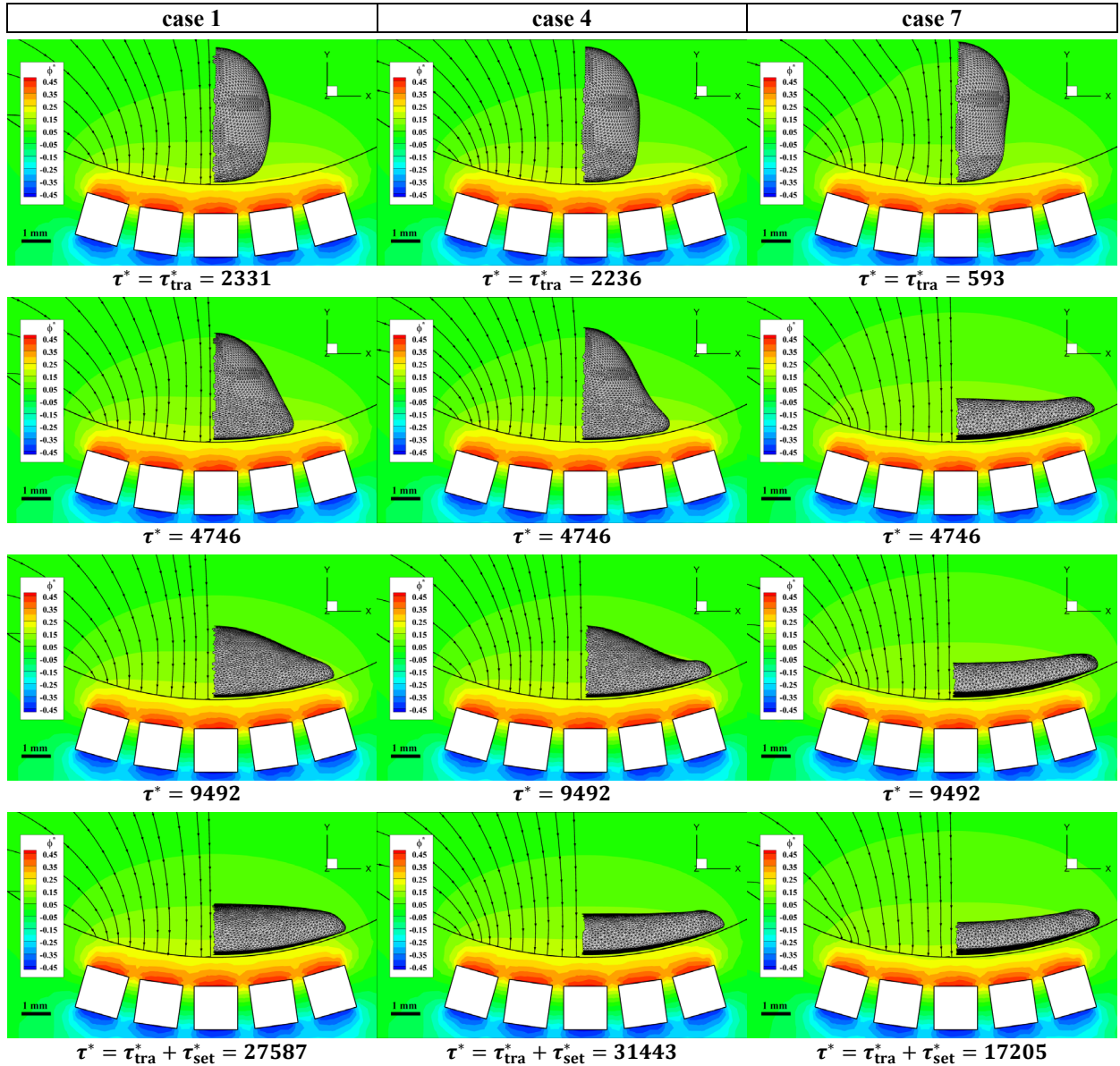

**Figure 14** The side view of the magnetic field lines and the (dimensionless) magnetic potential contours on the $z = 0$ plane along with a half of the drop front at different times, for case 1 ($\text{Bo}_m = 8542$, $\mu_d/\mu_c = 1.6$), 4 ($\text{Bo}_m = 21355$, $\mu_d/\mu_c = 1.6$), and 7 ($\text{Bo}_m = 21355$, $\mu_d/\mu_c = 4.0$).

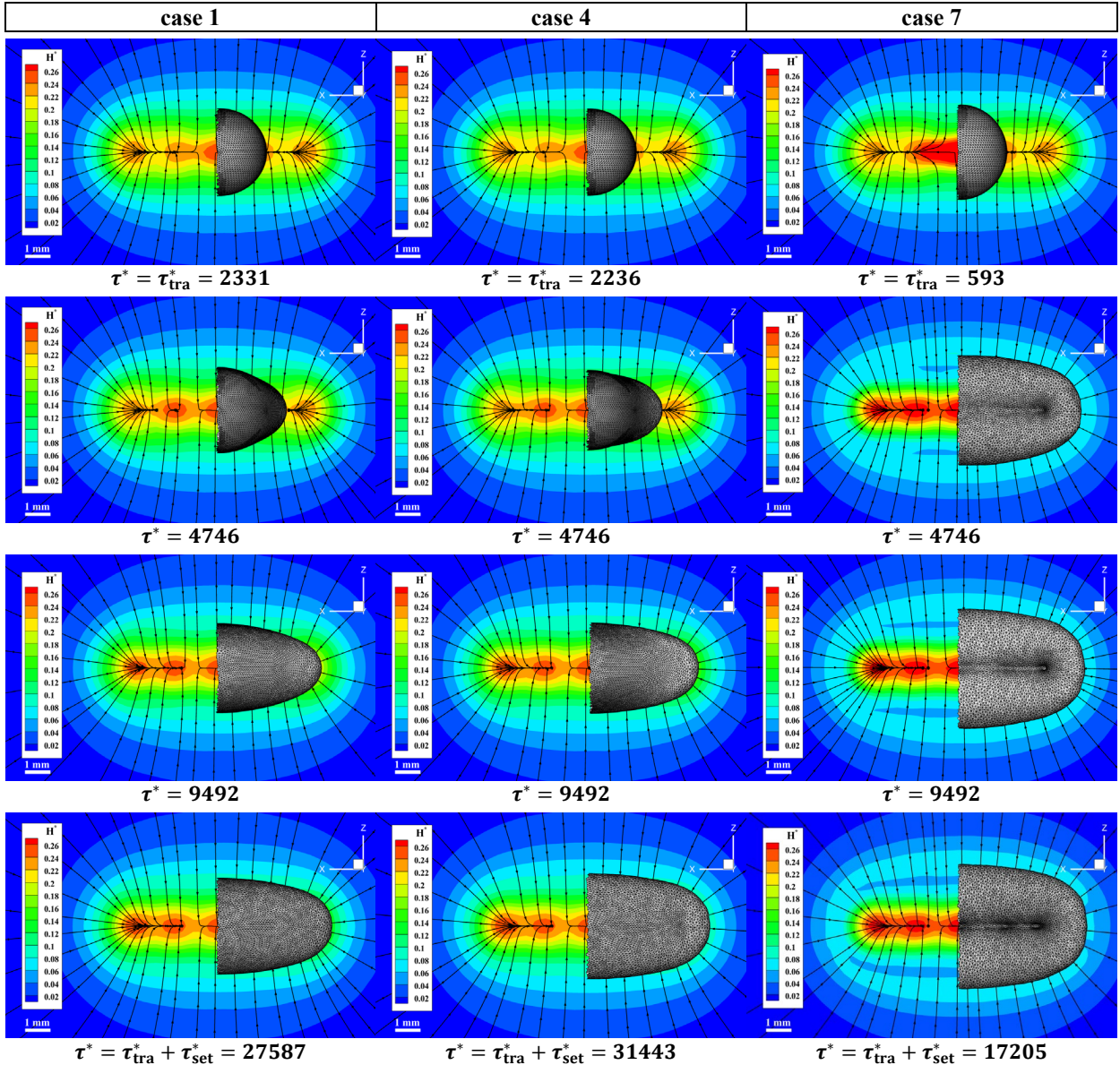

**Figure 15** The top view of the magnetic field lines and the (dimensionless) magnetic field magnitude contours on the eye surface along with a half of the drop front at different times, for case 1 ($Bo_m = 8542$, $\mu_d/\mu_c = 1.6$), 4 ($Bo_m = 21355$, $\mu_d/\mu_c = 1.6$), and 7 ($Bo_m = 21355$, $\mu_d/\mu_c = 4.0$).

The reason behind this phenomenon is that at large $\mu_d/\mu_c$, the presence of the drop significantly alters the external magnetic fields imposed by the magnets. The discontinuous magnetic field lines at the location of the drop surface for case 7 in figure 14 clearly shows this. Comparing the contours of $H^*$ at the final time for all cases in figure 15, it can be seen that the magnetic field alteration in case 7 leads to a larger magnetic field intensity near the retinal surface just above the magnets. This stronger magnetic field exerts a larger force on the drop surface which fiercely pulls a portion of the upper drop surface above the magnets towards the wall and creates the crease on the top surface. Approximately, at the same time when this crease is formed, the recoil-like behavior, observed in figure 13b, occurs which suggests that the recoil is originated from the generation of this crease. The transition of the shape evolution from the fast to the slow modes occurs at about $\tau^* = 9492$ for case 1 and 4 ($Bo_m = 8542$ and $21355$ in figure 13a) and at about $\tau^* = 4746$ for case 7 ($Bo_m = 21355$ in figure 13b). As observed in figure 14, at this time, the leading edge of the spreading droplet passes above the last magnet. Beyond that the magnetic field declines abruptly at the location of the leading edge and the expanding magnetic force decays. The reason for the appearance of the slow mode in the droplet spreading and retinal surface coverage, and the arrangement and location of magnets, predominantly determines at what point of spreading this mode occurs.

To analyze the stress and force exerted on the retinal surface, the contours of the dimensionless FDT compressive stress, $p^*_{FDT}$, on the eye wall for different cases, are compared in figure 16 along with the integral of this compressive stress on an area containing the drop, $F^*_{FDT} = \int_A p^*_{FDT}\, dA^*$, in figure 17. Note that the pressure used in the calculation of the FDT stress, $p_{FDT} = p - \tau_{nn}$, is the relative pressure with respect to the lowest pressure value on the retinal surface. According to figure 16, the greatest total FDT stress is concentrated on the retinal area just above the magnets' locations. As $Bo_m$ increases at a constant magnetic force, by decreasing the surface tension, the total computed magnetic force on the retinal surface is unchanged (figure 17) while the retinal coverage area by the drop increases (figure 13a). Therefore, the critical retinal areas of the largest total FDT stress shrink, as seen by comparing figure 16a and b. As a result, a lower drop-VH surface tension can produce a more uniform stress distribution on the retina at a constant effective FDT force. Using different configurations of magnets or magnets of different shapes can also control the stress concentration on the retina. On the other hand, when $Bo_m$ and $\mu_d/\mu_c$ increase simultaneously for a constant surface tension, the total magnetic force on the retinal surface rises (figure 17) and at the same time the coverage area expands significantly (figure 13a). However, the stress concentration at the retina still increases, comparing figure 16a and c, in spite of the increase in the coverage area. For case 7, $Bo_m$ and $\mu_d/\mu_c$ are 2.5 times those of case 1 (see table 4), while the peak FDT stress is approximately 4 times that of case 1 (figure 16). Therefore, when both $Bo_m$ and $\mu_d/\mu_c$ increase, avoiding excessive stress concentration is an important constraint to prevent retinal damage. Note that the stress concentration could also be controlled using different magnet shapes and arrangements.

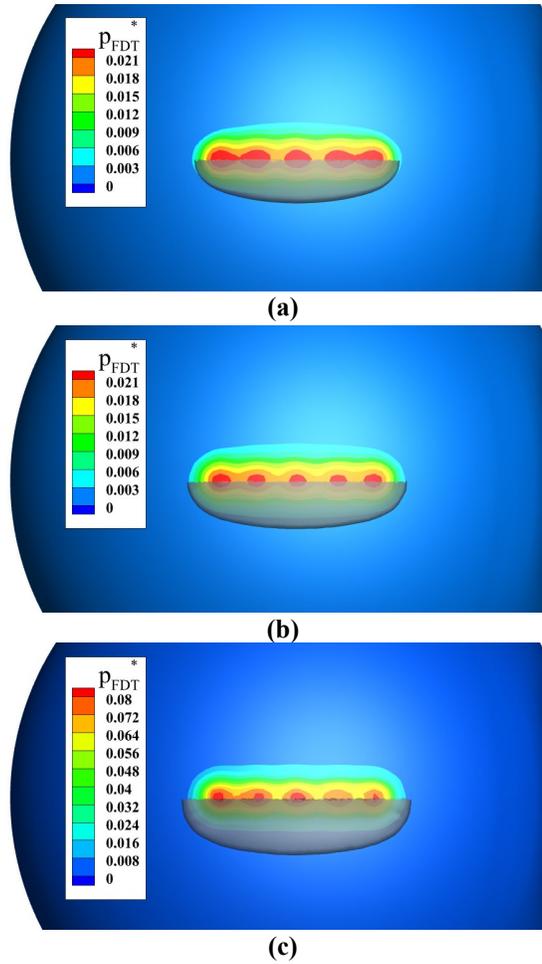

**Figure 16** The contours of $p^*_{FDT}$ on the eye surface and the bottom view of (a half of) the drop at $\tau^* = \tau^*_{set}$ for a) case 1, b) case 4, and c) case 7.

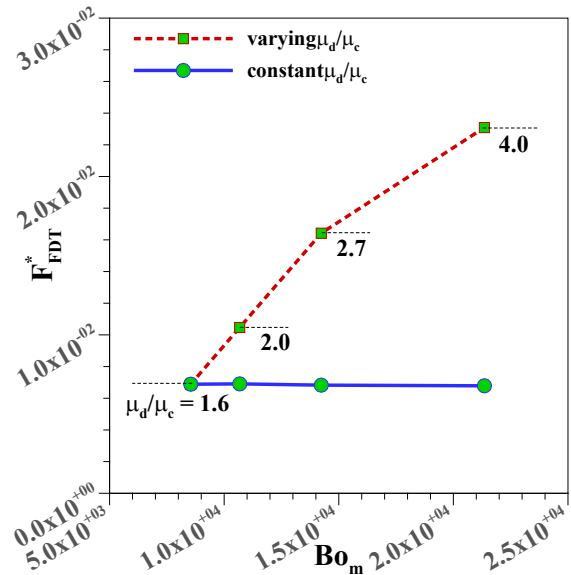

**Figure 17** The (dimensionless) total FDT force on the retinal surface versus the magnetic Bond number (a) at a constant $\mu_d/\mu_c$ (cases 1-4) and (b) by varying $\mu_d/\mu_c$ (cases 1, 5-7).

## 4.3. Single-magnet configurations

To study the effect of the magnet configuration, a simple scenario of a single magnet is used in this section. To be consistent with the situation considered in the previous works [5], the shape of the magnet is assumed to be a cylinder, triggering a 2D axisymmetric droplet shape change, with a radius of 1.5 *mm* and height of 1.5 *mm* instead of the cubic magnets in figure 1. To examine the effect of the fluid viscoelasticity, a simplified case of a Newtonian VH with a dynamic viscosity equal to the total viscosity, Eq. (25), of the realistic viscoelastic VH, assumed in the previous section, is also taken into consideration. The single-magnet cases, cases S1, N1, S2, N2, are defined in table 5, and the results of their simulation are compared with case 1, the multi-magnet configuration from the previous section. Note that all dimensionless parameters, except the geometrical ones, of case S1 are identical to those of case 1. For cases S2 and N2, the potential difference, $\Delta\phi_0$, of the magnet is doubled, compared to the other cases of this study.

**Table 5** The specification of cases with a single-magnetic configuration.

| Case | VH rheology | Magnet configuration | $\mu_d/\mu_c$ | La ($\times 10^5$) | $La_m$ ($\times 10^2$) | $Bo_m$ ($\times 10^{-3}$) | $\tau^*_{tra}$ | $\tau^*_{set}$ | $F^*_{FDT}$ ($\times 10^3$) |
|---|---|---|---|---|---|---|---|---|---|
| 1 | Viscoelastic | 5 cubic magnets | 1.6 | 1.11 | 9.51 | 8.542 | 2331 | 25255 | 6.9 |
| S1 | Viscoelastic | 1 cylindrical magnet | 1.6 | 1.11 | 9.51 | 8.542 | 4803 | 21004 | 1.5 |
| N1 | Newtonian | 1 cylindrical magnet | 1.6 | 1.11 | 9.51 | 8.542 | 14113 | 18220 | 0.53 |
| S2 | Viscoelastic | 1 cylindrical magnet | 1.6 | 1.11 | 38.05 | 34.17 | 981 | 16105 | 4.3 |
| N2 | Newtonian | 1 cylindrical magnet | 1.6 | 1.11 | 38.05 | 34.17 | 3150 | 25327 | 3.7 |

For the different cases in table 5, the values of the travel and settling time are reported, and the retinal coverage area variation over time is provided in figure 18. The comparison of cases 1 and S1 reveals the strong effect of the magnet configuration and magnetic field distribution on the objective parameters. The travel time increases (by a factor of about 2) and the coverage area reduces (by a factor of 1/4) using the single-magnet configuration, while the settling time reduces due to the less deformed droplet in this case. The droplet shapes, in figure 18 (right), are axisymmetric compared to the 3D elongated drops observed in figure 14 and figure 15. Comparing case N1 with S1 or N2 with S2 shows the important differences in the objective parameters predicted assuming a Newtonian medium instead of the realistic viscoelastic VH. The drop travel time in the Newtonian medium is significantly longer than that in the viscoelastic one. This is attributed to the strong shear-thinning effect of the VH, which is induced by the large mobility factor and polymeric-to-solvent viscosity ratio of the VH. Both these parameters decrease the apparent viscosity of a Giesekus fluid [57] near the drop surface and accelerate the drop motion and deformation. This phenomenon justifies the much larger initial rate of change of the coverage area for the viscoelastic cases compared to their Newtonian counterparts in figure 18 (left). This figure also indicates that the terminal coverage area in the viscoelastic medium is slightly larger than that in the Newtonian medium. This is due to the fact that in the stationary state, the viscous stress diminishes in both media while the viscoelastic medium induces an additional polymeric stress due to its elastic nature and the memory effect. This extra stress also increases the FDT compressive stress, $p^*_{FDT}$, concentration on the eye wall for the viscoelastic medium, comparing S2 and N2 in figure 18 (right), as well as the total FDT force on the retinal surface (see table 5).

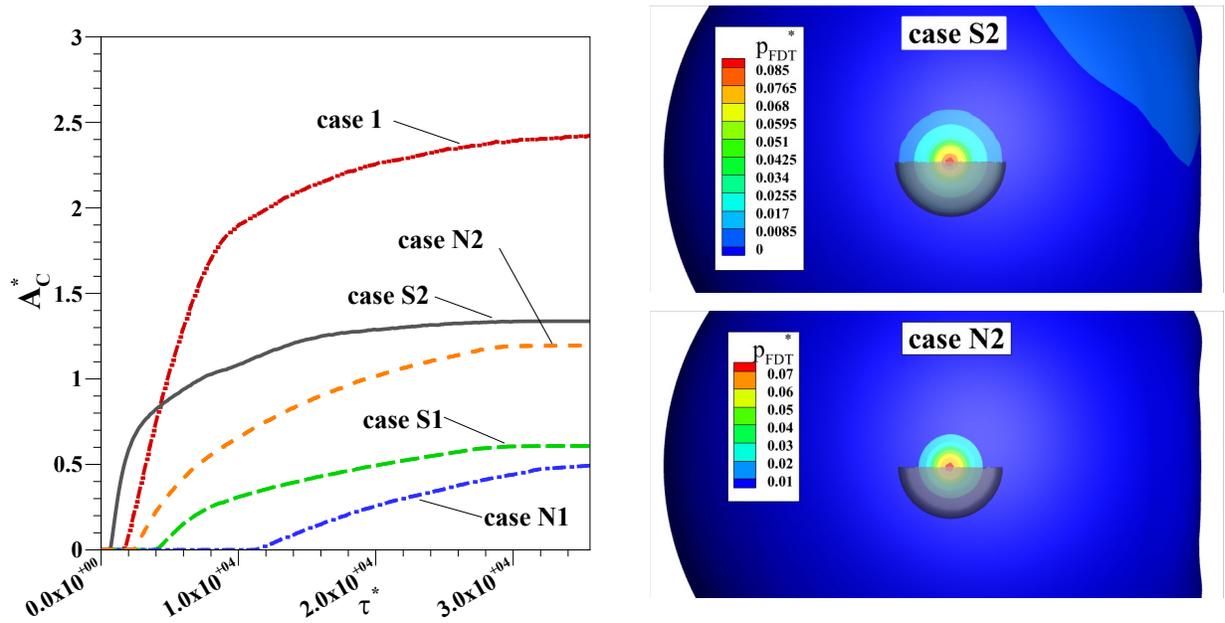

**Figure 18** The (dimensionless) retinal coverage area ($A^*_C$) over time (left) and the terminal contours of $p^*_{FDT}$ on the eye surface (right) for different case studies in table 5.

## 5. Conclusion

In the present study, the RD treatment via FDT in a real 3D eye geometry and magnets arrangement considering the viscoelastic rheology of the VH was studied numerically, for the first time. First, an FTM strategy for general 3D unstructured Eulerian grids was introduced. It was shown that for a robust and accurate FTM application to an FDT problem, intense noises on the front and large volume conservation errors induced by the near-wall effects should be avoided. To address the former issue, different front smoothing algorithms, including TSUR3D, VCS III, and VCS IV were tested and it was concluded that VCS IV with the frequency of 40 performs well. The latter challenge was also tackled by implementing a modified volume correction algorithm. The numerical model was then validated against three canonical benchmarks to assess the interface tracking, FHD, and viscoelastic sub-models, separately, avoiding the error-hiding effect. After that, the model was used to study the FDT-RD problem. The results demonstrated that the drop travel and settling times decline with an increase in the magnetic Laplace number, $\text{La}_m$, eventually reaching saturation at large $\text{La}_m$ values. On the other hand, as the Laplace number, La, grew, the settling time was reduced. The drop deformation increased by an increase in $\text{La}_m$ or decrease in La, while the rate of shape change was only a function of $\text{La}_m$. The deformation also showed a transition from a fast mode to a slow mode, in which the transition point depended on the magnet configuration and $\text{La}_m$ but on La. Most importantly, the terminal shape metrics, such as the retinal coverage area, depended on the ratio of the drop-to-VH permeabilities in addition to the magnetic Bond number, since this ratio can significantly alter the external magnetic field and the local magnetic force experienced by the drop. Increasing the Bond number and the permeability ratio significantly increased the total FDT force, retinal coverage area, and stress concentration; while increasing the Bond number thorough decreasing the surface tension at a constant magnetic force produced a more uniform FDT stress on the retina. It is worth noting that the shape of the magnets and their arrangement can also control the stress concentration. A myriad of multi-magnet configurations can be designed and investigated for the current problem, which is an interesting topic for future studies.

**Supplementary material:** The supplementary material provides an animation of FDT-RD treatment (case 4 from table 4).

**Declaration of Interests:** The authors report no conflict of interest

**Acknowledgments**

The authors would like to express their appreciation to Dr. Afsaneh Amani, Ophthalmologist, Fellow of the European Board of the Ophthalmology (FEBO), 40764 Langenfeld, Germany, for her valuable advice on ophthalmological aspects of the present research.

**Appendix A: The volume correction algorithm**
The present volume correction algorithm is performed by computing the corrected location of each front node by [47, 58]:

$$x_p^c = x_p + \epsilon n_p, \tag{30}$$

where $n_p$ is the local front unit normal vector at each node and the correction displacement magnitude, $\epsilon$, is determined by an optimization procedure [47, 58]. For our main problem, involving front surface near wall boundaries, we modified the original algorithm slightly and applied Eq. (30) only to front nodes which are not on the drop-wall contact or coverage area ($A_c$), i.e., not on the threshold surface. The parameter $\epsilon$ is then calculated by the solution of the following cubic equation (choose the $\epsilon > 0$ root if $V < V_0$ and $\epsilon < 0$ otherwise):

$$(V_0 - V) = a\epsilon^3 + b\epsilon^2 + c\epsilon, \tag{31}$$

where $V_0$ and $V$ are the initial and current volumes of the droplet, and the coefficients $a$, $b$, and $c$ are computed by:

$$a = \frac{1}{6} \sum_{\substack{m=1 \\ m \notin A_c}}^{N_E} [n_1 \cdot (n_2 \times n_3)]_m,$$

$$b = \frac{1}{6} \sum_{\substack{m=1 \\ m \notin A_c}}^{N_E} [x_1 \cdot (n_2 \times n_3) + x_2 \cdot (n_3 \times n_1) + x_3 \cdot (n_1 \times n_2)]_m, \tag{32}$$

$$c = \frac{1}{6} \sum_{\substack{m=1 \\ m \notin A_c}}^{N_E} [n_1 \cdot (x_2 \times x_3) + n_2 \cdot (x_3 \times x_1) + n_3 \cdot (x_1 \times x_2)]_m,$$

where the summations are over all front triangular elements, $m$, not located in the front-wall coverage region. $x_1$, $x_2$, and $x_3$ are the corner vertices of each element, and $n_1$, $n_2$, and $n_3$ the front unit normal vectors at these vertices. The Newton's method with a relative error tolerance of $10^{-7}$ is used to solve Eq. (31). The volume correction step is applied when $|V - V_0|/V_0 > 0.001$.

### Appendix B: Validation of the FHD model

In the second benchmark illustrated in figure 19a, an initially spherical ferrofluid droplet is suspended in a carrier fluid and subjected to a uniform constant external magnetic field of intensity $H_0$ in the z-direction. For negligible gravity, Afkhami, et al. [33] reported the equilibrium deformation of the droplet experimentally. The deformation is quantified by the ratio of the major-to-minor axes of the deformed droplet, i.e., $b/a$ in Figure 19a. They also provided an analytical solution for the small deformation limit as a function of the magnetic Bond number, $Bo_m$, and the permeability ratio, $\mu_c/\mu_d$. For the present simulations, a spherical droplet of radius $R_d = 1.29\ mm$ is placed at the center of a cylindrical domain of diameter $D = 12R_d$. The physical properties of the fluids are extracted from the experiment and given in table 6. The magnetic susceptibility of the droplet is constant and equal to $\chi_d = 0.8903$ under low external magnetic fields ($H_0 < 7\ kA/m$) and varies as $\chi_d = (4.8568 \ln H_0 - 3.956)/H_0$ for higher magnetic field intensities ($H_0 \geq 7\ kA/m$).

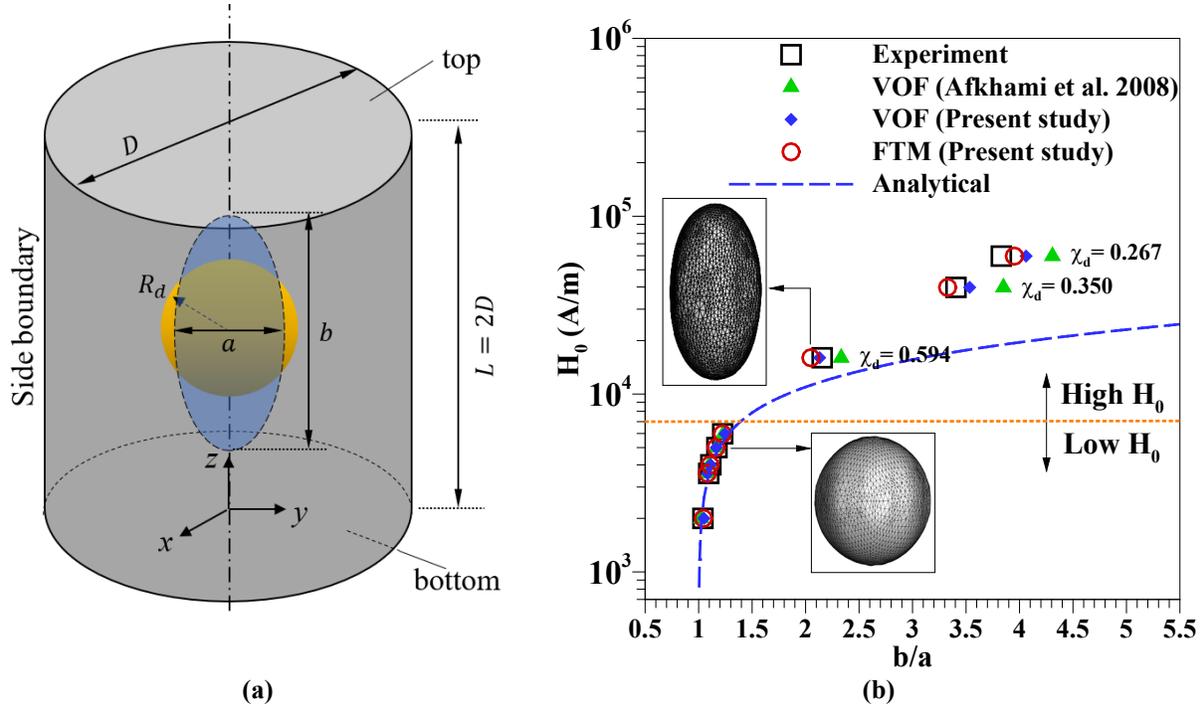

(a) (b)

**Figure 19** Benchmark 2: a) The schematic geometry and computational domain. The equilibrium shape of the droplet is indicated by the semi-transparent prolate spheroid. b) The comparison of the present simulations with the experimental, numerical, and analytical results by Afkhami, et al. [33].

**Table 6** Benchmark 2: The physical properties of drop and carrier phases [33].

| $\rho_c\ (kg/m^3)$ | $\rho_d\ (kg/m^3)$ | $\eta_c\ (Pa.s)$ | $\eta_d\ (Pa.s)$ | $\sigma\ (N/m)$ | $\chi_c$ | $\chi_d$ |
|---|---|---|---|---|---|---|
| 1400 | 1260 | 0.1 | 0.1 | 0.0135 | 1.0 | variable |

The no-slip boundary condition is applied to all cylindrical domain boundaries. For the magnetic field potential, a zero-gradient condition is applied at the side boundary, zero-value potential at the bottom, and a fixed-value potential $\varphi_0$ at the top boundary, where $\varphi_0$ is determined based on the imposed external magnetic field intensity as: $\varphi_0 = H_0 L$. A uniform block-structured O-type grid, similar to figure 3b, with at least 40 CPD and total 367200 cells was found to be required to achieve grid-independent results. Simulations for different intensities of the magnetic field in the range $2.0 \leq H_0 \leq 59.67\ (kA/m)$ have been conducted and the results are compared against the reference numerical results, empirical data, and analytical solution for the low-deformation limit in figure 19b. For the sake of comparison, we performed a VOF-MULES simulation, with a solver detailed in reference [59], in addition to FTM. For low $H_0$ (low-deformation limit), all results coincide with a negligible difference. The model predictability at high $H_0$ values is particularly of interest here since the main problem (section 4.2) falls within the high-deformation range. For large values of $H_0$, where the drop undergoes large deformations, the present FTM results in the best agreement with the empirical data. The deviation between VOF-PLIC results by Afkhami, et al. [33] and the present VOF-MULES is likely due to the details of the numerical methods, most importantly the interface capturing strategy, i.e., MULES versus PLIC.

**Appendix C: Validation of the viscoelastic model**

The third benchmark considers the flow of a viscoelastic fluid through a planar contraction studied experimentally by Quinzani, et al. [60]. They provided a standard database which has been widely-used to validate viscoelastic solvers, e.g., [41, 61]. We chose case 3 from this database which possesses a very similar rheological behavior to VH and its constitutive law is described by a 4-mode Giesekus model with the parameters given in table 7. The fluid density is 803.87 $kg/m^3$ and outlet bulk velocity equals $U_{out} = (H/h)U_{in} = 7.44 \, cm/s$. The geometry of the contraction is shown in the inset of figure 20. The dimensionless governing parameters are:

$$\text{Re} = \frac{2\rho U_{out} h}{\eta_0}, \frac{H}{h}, \text{De}_n = \lambda_n \frac{U_{out}}{h}, \alpha_n, \frac{\eta_{p,n}}{\eta_s} \, . \tag{24}$$

where for case 3, $\text{Re} = 0.27$, $H/h = 3.97$, and $\text{De}_n$ is reported for each mode in table 7.

**Table 7** Benchmark 3: The 4-mode Giesekus model parameters [60]. The solvent viscosity is $\eta_s = 0.002 \, \text{Pa.} s$.

| Mode | $\alpha_n$ | $\lambda_n$ (s) | $\eta_{p,n}$ (Pa.s) | $\text{De}_n$ |
|---|---|---|---|---|
| 1 | 0.5 | 0.6855 | 0.0400 | 15.94 |
| 2 | 0.2 | 0.1396 | 0.2324 | 3.25 |
| 3 | 0.3 | 0.0389 | 0.5664 | 0.90 |
| 4 | 0.2 | 0.0059 | 0.5850 | 0.14 |

The boundary conditions are the uniform value of $U_{in}$ for the velocity and zero-value for the polymeric stress at the inlet, no-slip condition for the velocity at the wall, zero-value for the y-velocity at the symmetry plane, zero-value for the pressure at the outlet, and zero gradients for the rest of the boundary conditions. The present numerical results are compared with the experimental data [60] in figure 20. The present results are in fine agreement with the experiment which shows the validity of our computational model.

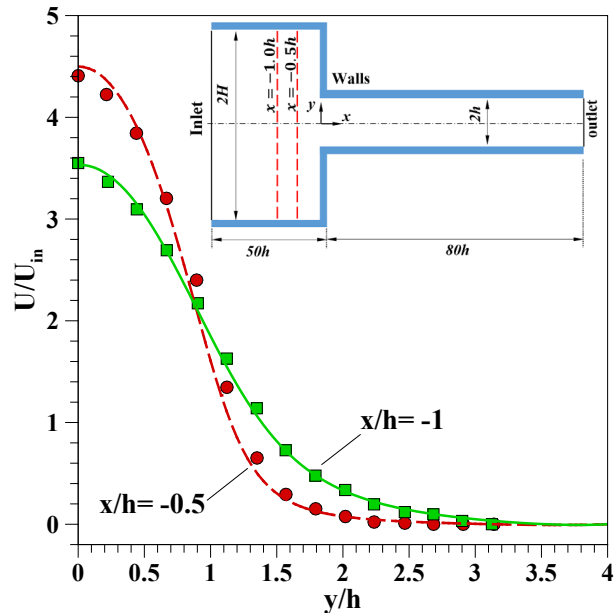

**Figure 20** Benchmark 3: The comparison of the present simulation (lines) and experimental data [60] (symbols) at two different cross-sections. The inset shows the schematic geometry and computational domain ($h = 3.2\ mm$).